\documentclass[review]{elsarticle}
\usepackage{lineno,hyperref}
\usepackage[ruled,vlined]{algorithm2e}
\usepackage{amsmath}
\usepackage{booktabs}
\usepackage{longtable}
\usepackage{graphicx}
\usepackage{url}
\usepackage{multicol}
\usepackage{caption}
\usepackage{subcaption}
\usepackage{amsmath,amssymb}
\usepackage{enumitem}
\usepackage{tikz}
\usepackage{verbatim}
\usepackage{hyperref}
\usepackage{multirow}
\usepackage{pgfplots}
\usepackage{listings}
\usepackage{tabularx}
\usepackage{wrapfig}
\usepackage{xcolor}
\usepackage{float}

\usepackage{color, colortbl}
\definecolor{Gray}{gray}{0.95}

\setcitestyle{square}

\definecolor{codegreen}{rgb}{0,0.6,0}
\definecolor{codegray}{rgb}{0.5,0.5,0.5}
\definecolor{codepurple}{rgb}{0.58,0,0.82}
\definecolor{thesisblue}{RGB}{38,111,126}

\lstdefinestyle{mystyle}{
   backgroundcolor=\color{Gray},
   commentstyle=\color{codegray},
   keywordstyle=\color{thesisblue},
   numberstyle=\tiny\color{codegray},
   stringstyle=\color{codegray},
   basicstyle=\footnotesize,
   breakatwhitespace=false,
   breaklines=true,
   captionpos=b,
   keepspaces=true,
   numbers=left,
   numbersep=5pt,
   showspaces=false,
   showstringspaces=false,
   showtabs=false,
   tabsize=2
}

\usetikzlibrary{calc, automata, positioning, shadows, arrows, shapes.geometric}

\pgfplotsset{compat=1.8}
\makeatletter
\pgfarrowsdeclare{DIN}{DIN}
{
 \pgfutil@tempdima=0.5pt%
 \advance\pgfutil@tempdima by.25\pgflinewidth%
 \pgfutil@tempdimb=7.29\pgfutil@tempdima\advance\pgfutil@tempdimb by.5\pgflinewidth%
 \pgfarrowsleftextend{+-\pgfutil@tempdimb}
 \pgfutil@tempdimb=.5\pgfutil@tempdima\advance\pgfutil@tempdimb by1.6\pgflinewidth%
 \pgfarrowsrightextend{+\pgfutil@tempdimb}
}
{
 \pgfutil@tempdima=0.5pt%
 \advance\pgfutil@tempdima by.25\pgflinewidth%
 \pgfsetdash{}{+0pt}
 \pgfsetmiterjoin
 \pgfpathmoveto{\pgfpointadd{\pgfqpoint{0.5\pgfutil@tempdima}{0pt}}{\pgfqpoint{-4mm}{0.5mm}}}
 \pgfpathlineto{\pgfqpoint{0.5\pgfutil@tempdima}{0\pgfutil@tempdima}}
 \pgfpathlineto{\pgfpointadd{\pgfqpoint{0.5\pgfutil@tempdima}{0pt}}{\pgfqpoint{-4mm}{-0.5mm}}}
 \pgfpathclose
 \pgfusepathqfillstroke
}
\pgfarrowsdeclarereversed{DIN reversed}{DIN reversed}{DIN}{DIN}
\tikzstyle{startstop} = [rectangle, rounded corners, minimum width=3cm, minimum height=1cm,text centered, draw=black, fill=red!30, text width=3cm]
\tikzstyle{io} = [trapezium, trapezium left angle=70, trapezium right angle=110, minimum width=3cm, minimum height=1cm, text centered, draw=black, fill=blue!30]
\tikzstyle{process} = [rectangle, minimum width=1.8cm, minimum height=1cm, text centered, text width=2.4cm, draw=black, fill=orange!30]
\tikzstyle{decision} = [diamond, minimum width=2cm, minimum height=1cm, text centered, draw=black, text width=2cm, fill=green!30]
\tikzstyle{decisionspecial} = [diamond, aspect=2, minimum width=2cm, minimum height=1cm, text centered, draw=black, text width=5cm, fill=green!30]

\tikzstyle{state} = [rectangle, minimum width=4cm, minimum height=1.2cm, text centered,  draw=gray!30, fill=gray!30]
\tikzstyle{arrow} = [thick,->,>=stealth, line width=0.5mm]
\tikzstyle{greenarrow} = [thick, ->, >=stealth, draw=green!50!black, line width=0.5mm]
\tikzstyle{redarrow} = [thick, ->, >=stealth, draw=red, line width=0.5mm]

\hypersetup{allcolors=[RGB]{31,93,105}, colorlinks=true}

\modulolinenumbers[5]

\begin{document}

\begin{frontmatter}

\title{SEVA: A Data driven model of Electric Vehicle Charging Behavior.}

\author[mymainaddress,mysecondaryaddress]{Jurjen R. Helmus}\corref{mycorrespondingauthor}
\cortext[mycorrespondingauthor]{Corresponding author}
\ead[url]{www.idolaad.nl}

\author[mysecondaryaddress]{Seyla Wachlin}
\author[mysecondaryaddress]{Igna Vermeulen}
\author[mysecondaryaddress]{Mike H. Lees}

\address[mymainaddress]{University of Applies Sciences Amsterdam, weesperzijde 190 1091DZ, Amsterdam}
\address[mysecondaryaddress]{UvA Computational Science Lab, Science park ,Amsterdam }
\address[mtertiaryaddress]{ITMO something in Russia}

\begin{abstract}
Governments and cities around the world are currently facing rapid growth in the use of Electric Vehicles and therewith the need for Charging Infrastructure. 
For these cities, the struggle remains how to further roll out charging infrastructure in the most efficient way, both in terms of cost and use. 
Forecasting models are not able to predict more long-term developments, and as such more complex simulation models offer opportunities to simulate various scenarios.
Agent based simulation models provide insight into the effects of incentives and roll-out strategies before they are implemented in practice and thus allow for scenario testing. 
This paper describes the build up of an agent based model that enables policy makers to anticipate on charging infrastructure development. 
The model is able to simulate charging transactions of individual users and is both calibrated and validated using a dataset of charging transactions from the public charging infrastructure of the four largest cities in the Netherlands.

\end{abstract}

\begin{keyword}
\texttt{agent based simulation, charging behavior, charging infrastructure}\sep  \sep template
%\MSC[2010] 00-01\sep  99-00 % no clue what this is
\end{keyword}

\end{frontmatter}

%\linenumbers

\section{Introduction}

Governments and cities around the world are currently facing rapid growth in the use of Electric Vehicles (EVs). Many countries identify the need for supporting this growth through the proper deployment of public charging infrastructure. Potential EV users may hesitate when purchasing an EV if there is insufficient charging infrastructure, while governments are hesitant to provide new charge poles (CPs) if the number of EV users is limited. Nonetheless, cities are creating and expanding their charging infrastructure to facilitate demand and to increase EV adoption~\citep{Leurent2011,Spoelstra2014, Franke2013, Helmus2018a}. Cities that have started with charging infrastructure roll-out since early market development, now have more mature and city-wide charging infrastructures \cite{VanDenHoed2014}.
For these cities, the struggle remains how to further roll out charging infrastructure in the most efficient way, both in terms of cost and use. Both over-capacity and under-capacity of the CPs are situations which governments try to avoid. Predicting the future usage of CPs for a current, or future, population of EV users would enable policy makers to create a more efficient charging infrastructure~\citep{Helmus2015c}.

% Problem Statement
Predictive simulation models provide insight into the effects of incentives and roll-out strategies before they are implemented in practice and thus allow for scenario testing. While simulation models on this topic exist, these are, to the best of our knowledge, not validated or only validated using small amounts of data~\citep{Hess2012a, Sweda2011b, Zhu2016, Uhrig2015, Xi2013b, Yi2016, Momtazpour2012a, Paffumi2015a}. This greatly decreases the predictive certainty of the models, which is identified as a limitation in both~\citep{Sweda2011b} and~\citep{Momtazpour2012a}.

In this paper we describe SEVA, a data-driven model of Electric Vehicle (EV) charging behavior. This model is able to simulate charging transactions of individual users and is both calibrated and validated using a dataset of charging transactions from the public charging infrastructure of the four largest cities in the Netherlands as described in ~\cite{VanDenHoed2014}. This is one of the largest EV transaction datasets available for research \footnote{The dataset is exclusively available for the ido-laad project.} , with over 5.4 million transaction dating from 2014 to 2017. The main purpose of the SEVA model is to provide insights into the effects of incentives and roll-out strategies by providing scenario testing.

The remainder of this paper is organized as follows. In the Section~\ref{sec:2lit_review}, we present an overview of relevant literature, from which we conclude what aspects of the EV system should be covered by the model. Thereafter, in section \ref{sec:model description} a detailed description of the model is given according to the ODD protocol ~\cite{Grimm2006,Muller2014}. We then examine output metrics of the model to validate and test the model in Section \ref{sec:metrics}. In section \ref{sec:base_model_evaluation} we show an sensitivity analysis of the model parameters and present the validation of the model both in terms of the behavior of EV users and the use of CPs in the model. The paper ends with a conclusion and discussion on the applicability of the SEVA model.

%%%% ---------------------------- %%%%
%%%% ---------------------------- %%%%
%%%% Review of Related Literature %%%%
%%%% ---------------------------- %%%%
%%%% ---------------------------- %%%%

\section{Review of Relevant Literature}
\label{sec:2lit_review}

Various papers in the field of simulation and modeling study EV charging infrastructure, each with its own focus. Sweda and Klabjan \cite{Sweda2011b} and Hess et al \cite{Hess2012a} are concerned with minimizing charging costs, Zhu \cite{Zhu2016} focuses on improving the adoption of EVs, and~\cite{Uhrig2015, Xi2013b, Yi2016, Paffumi2015a, Momtazpour2012a, Daina2017c} all focus on optimizing the energy demand. Besides models dedicated to charging infrastructure, there exist core traffic simulation models that extend to EV adoption and EV infrastructure~\cite{Djanatliev2017}.

Most research on simulation models build upon assumptions regarding the behavior of EV users, without the use of actual user data. Data based simulation models either use data from Internal Combustion Engine (ICE) vehicles from which EV behavior is extrapolated or use stated response data on EV infrastructure requirements \cite{Philipsen2015a}. The former contain biases which cannot be validated without available data of EV usage and the latter carry biases inherent to hypothetical choice situations~\citep{Daina2017c}. Data based models mainly contain road data, population data, parking data, stated response data or driving data of ICE vehicles are used. Several of the authors point out the lack of EV data as a limitation~\citep{Sweda2011b,Momtazpour2012a}. To the best of our knowledge, currently no EV simulation model exists that is validated using large amounts of real-world data about EV usage (i.e. revealed preference data).

The importance of specifically understanding EV user charging behavior is pointed out by Azadfar et al. \cite{Azadfar2015}, Franke et al. \cite{Franke2013} and Sweda et al. \cite{Sweda2011b}. Sweda et al. \cite{Sweda2011b} states that understanding and implementing charging behavior in a simulation model is essential to optimize the charging infrastructure roll-out and to promote a more efficient utilization of the infrastructure. In order to scope the dynamics of this model it is necessary to define charging behavior. According to Helmus et al. \cite{Helmus2015c} charging behavior can be defined as the successful result of the intentional behavior to charge a specific EV at a specific CP for a specific duration. This could define charging behavior to some extent, but it is likely that more dimensions of charging behavior need to be considered to fully capture the complex behavior of EV users in a model. De Gennaro et al. \cite{DeGennaro2015} define the charging behavior of an EV user as a portfolio of charging transactions based on charging strategies. While these charging strategies may be difficult to implement, the concept of a portfolio of charging transactions reflects the data based approach.

Most charging behavior studies focus on either the psychological side of charging behavior, making use of inquiries, interviews and stated response experiments~\citep{Franke2013, Daina2017c}, or the effect of EVs on the power grid~\citep{Kintner-Meyer2007}. The former look at behavior as a decision making process, while the latter capture behavior as the activity profiles of EV users~\citep{Helmus2015c}. Helmus et al \citep{Helmus2015c} state that the mean behavior of EV users might be a bad estimator of real behavior. An example of this is an EV user who mainly charges at one of two charge times (8am or 6pm), but never at the mean of those two times (2pm).

Next to research on activity perspective of charging behavior, there is literature that identified the factors that influence the behavior itself. An important study by Franke et al. \cite{Franke2013} shows the user-battery interaction style as the main variable for EV user behavior. This variable provides a balance between how much users are influenced by either their state of charge or the time between transactions. With a low user-battery interaction style the time between transactions is the key factor and state of charge is less important. Spoelstra \cite{Spoelstra2014} performed a literature review to address the various factors which influence charging behavior and distinct three categories, namely driver related factors (range anxiety, planning, mobility pattern and EV experience), vehicle related factors (battery size, vehicle range and vehicle type) and environment related factors (CP density). Spoelstra \cite{Spoelstra2014} also states that charging behavior itself can be conceptualized by six dimensions, namely CP location, CP type, charging frequency, time of day, charging duration and energy transfer. Azadfar et al. \cite{Azadfar2015} also investigated the factors that influence charging behavior, and the dimensions of this charging behavior. They consider the time of day, the duration of the charging transaction, the frequency of charging and the energy required to charge the vehicle batteries. They state that factors which influence this behavior are the EV penetration rate, the charging infrastructure, the battery performance of EVs, the costs and incentive programs. They identify two key factors which most strongly influence charging behavior, namely the charging infrastructure (environment related) and the battery performance (vehicle related).

This research builds upon the real charging transactions of each EV user, which implies that the vehicle and infrastructure related factors influencing the behavior are implicitly taken into account in the model. More specifically, this study focuses on EV user activity patterns and their effect on charging infrastructure requirements. This implies that transaction volume (in kWh) is considered out of scope, even though the impact of charging may be of interest. The dimensions of charging behavior used in this research time of charging, location of charging and duration of charging. We leave out the charging frequency factor, simply because this can be inferred from the charging duration and inter-arrival duration.

%%%% ---------------------------- %%%%
%%%% ---------------------------- %%%%
%%%% Model Description            %%%%
%%%% ---------------------------- %%%%
%%%% ---------------------------- %%%%

\clearpage
\section{Model Description}
\label{sec:model description}
This section presents a description of the SEVA model. In the following subsections we describe workings of this model, following ODD+D \cite{Muller2013}, an extended version of the ODD (Overview, Design concepts, Details) protocol~\citep{Grimm2006}.

\subsection{Overview: Purpose}
The main purpose of the SEVA model is to provide insights into the dynamics of interacting EV users and future roll-out scenarios of charging infrastructure. The model is developed for policy makers to allow for decision making optimization by scenario testing. As such, it is important that the simulation results mimic the real system as exact as precisely. Therefore, each agent has a unique behavioral pattern that matches its self generated real-world data. % each agent has a unique behavioral pattern based on its data
The data itself is based on charging sessions from different location types (e.g. residential and office areas). Therefore, no assumptions are needed on the type of behavior related to the location type, such as traveling to home and work. This allows to run the model on any other type of location as long as the data is available.
%The proposed SEVA model is data driven to setup the model close to the real world system and validated against the real world data.
Dynamic settings for users base and infrastructure configuration allow for investigation of roll-out scenarios. The model allows to reveal dynamics not captured in the charging data such as competition for resources that led to failed connection attempts.

% % % % % % % % % % % % % % % % % % % % % % %
%            MODEL DESCRIPTION              %
% % % % % % % % % % % % % % % % % % % % % % %

\subsection{Overview: Entities, State Variables and Scales}
\label{sec:entities}
The model contains three entities, (1) the environment, (2) the agents and (3) the simulation handler. An overview of the entities and descriptions of their state variables can be found in in the appendix in table \ref{tab:entities}. The environment is defined as the collection of all CPs together with all the information about these CPs (i.e. their spatial location and their placement date). Each CP has two or more sockets and thus multiple agents can be connected to a single CP at the same time. The sockets can be in a state of occupied and available. If several agents are connected at a CP such that all sockets are taken, then no additional agent can connect to this CP. This results in a failed connection attempt if one extra agent tries to connect. 

Agents are defined as the combination of the EV users and their EV, as they are identified in the data by their charging card. This implicitly means that more users of the same vehicle (e.g. in families) are regarded as one agent. Agents in the system can be in one of three state variables: \textit{connected}, \textit{disconnected} or \textit{selecting CP}. Every agent is has complete knowledge of the environment, meaning it knows where CPs are located. The agents do not know whether an outlet of selected CP is occupied until it tries to connect to the CP. The only form of interaction between agents in the system is via the occupation of CPs. The agents do not communicate witch each other and as such do not share information on the state of other entities such as CPs in the system. The interaction between the users is limited to patterns that do not require information or communication. In this research, the exogenous factors that affect the simulation results are the increase in user base and and the growth of the infrastructure. 

The overall simulation is managed by a simulation handler. This manager controls the actions of the agents and also the interactions between the agents and the environment. Furthermore, this simulation handler has a collection of observers which keep track of metrics for the experiments and system evaluation, such as the average simulation time and the agent validation scores. The spatial scale of the model is continuous and all valid combinations of longitude and latitude within the boundaries of the Netherlands can be used. The model focuses on simulating EV use in the four major metropolitan areas of the Netherlands (G4 cities: Amsterdam, Rotterdam, the Hague and Utrecht). The temporal scale is discretized using bins of $T$ minutes. $T$ is an input parameter of the model and the value for this variable can be found in \ref{tab:parameters data}. The simulation can be run for a chosen number of days, months or years. The exact start date and stop condition that are used for the experiments and analysis are considered input parameters. They can be found in Appendix \ref{sec:model parameters} (Table~\ref{tab:parameters simulation}). Both the number of agents and CPs in the system can grow artificially with scenario planning, or can reflect the real growth of EV users in the Netherlands.

% % % % % % % % % % % % % % % % % % % % % % %
%            PROCESS OVERVIEW               %
% % % % % % % % % % % % % % % % % % % % % % %

\subsection{Overview: Process Overview and Scheduling}
\label{sec:process overview}
The simulation handler is a discrete event scheduler that controls the actions of the agents by storing all agents in a time ordered queue, where the time of an agent is the time of its next activity. The simulation handler sequentially pops agents from this queue. Once an agent is popped, it executes its next activity (either a connection or a disconnection activity) and then recalculates the time of its next activity. The simulation handler then pushes the agent back into the ordered queue.

As mentioned, agents in the system can be in one of three states, namely \textit{connected}, \textit{disconnected} or \textit{selecting CP}. The transitions between these states are controlled by the processes \textit{connection}, \textit{disconnection} and \textit{CP selection}. The execution loop connecting the states and processes can be seen in Fig.\ref{fig:agent execute activity}. When a connected agent executes its next activity, it will disconnect using the disconnection process. When an agent is disconnected, this agent will choose a CP from a set of CPs (which we call \emph{cluster}) using the connection process and then it will try to choose a CP to connect to by using the CP selection process. Table \ref{tab:process_overview} in section  \ref{subsection:process_overview} provides an overview of the processes. A detailed description on the processes can be found in section \ref{sec:submodels}.

% % % % % % % % % % % % % % % % % % % % % % %
%            DESIGN CONCEPTS                %
% % % % % % % % % % % % % % % % % % % % % % %

\label{TheoryBackground}
%general concept / theory / background
%While agent based models are common used tools for policy simulation, there are only a limited number of data-driven models available \cite{Thiele2014}. %reference to As such, there is limited theory available on the design concepts of data-driven ABMs.Research that does build upon data advises to use part of the data for calibration and validation of the model\cite{Zhang2016}.
In this research the dataset is split into a \emph{training} set and a \emph{validation} set and for each individual agent the results are validated against its actual data. %could be a bit too early  to state that here
The two datasets are created by setting a split date, a model parameter.
Splitting the data is preferred over sampling, since this would affect the behavioral distribution of disconnection time between two sessions.

For each agent present in the system, its behavior is generated as distributions in the \emph{time} and \emph{space} dimension based on its data in the training set. This implies that each agent has a unique set of distributions during simulation. The agent behavior within the model is not static, since (1) the  behavior in the simulation is sampled from the behavior distributions and (2) due to interaction between users, the agents may deviate from chosen locations. The distributions generated at initialization are updated accordingly during the simulation.

The agent learning based on interactions is out of scope of the current model, since this is implicitly present in the behavioral data. We assume that EV users display rational behavior in trying to optimize their convenience in charging as much as possible. We therefore assume that we approximate the behavior that the objective function would have displayed by the creation of rules from all historical charging transactions. The data does not provide the final destination of an agent, yet we assume that the charging data indicates a destination nearby.

\subsection{Design Concepts:Individual Decision-Making}
\label{IndividualDecisionMaking}
\begin{figure}[h]
    \center
    \begin{tikzpicture}[node distance = 3.5cm]
        \node (disconnected) [state] {\textbf{Disconnected}};
        \node (select) [state, right of=disconnected, yshift = -2cm, xshift = 1cm] {\textbf{Selecting CP}};
        \node (connected) [state, below of=disconnected, yshift = -0.5cm] {\textbf{Connected}};
        \draw[arrow] (disconnected) to[bend left=30]  node[midway,right] {\textit{Connection process}} node[midway,below] {} (select);
        \draw[arrow] (connected) to[bend left=40]  node[midway,left] {\textit{Disconnection process}} node[midway,above] {} (disconnected);
        \draw[greenarrow] (select) to[bend left=30]  node[midway,right] {\textcolor{green!50!black}{\textbf{Success}}} node[midway,above] {} (connected);
        \draw[redarrow] (select) to[bend left=30]  node[midway,left] {\textcolor{red}{\textbf{Failure}}} node[midway,above] {} (disconnected);
    \end{tikzpicture}
    \captionof{figure}{The activity loop of an agent. Depending on the state of the agent (the grey boxes) the different processes are called. Note that the colored lines indicate the two possible outcomes of the \textit{CP selection} process, namely success or failure.}
    \label{fig:agent execute activity}
\end{figure}
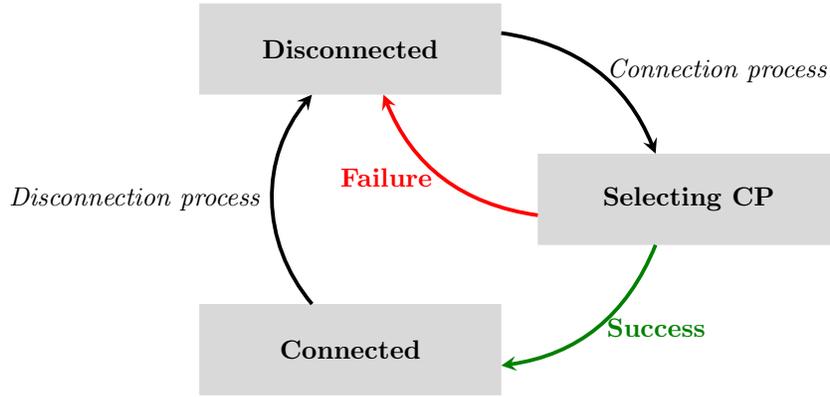

The subject of individual decision making includes the three processes disconnection, connection and CP selection, shown in \ref{fig:agent execute activity}. From the literature, it is known that EV users are habit-based and frequent only a limited number of CPs routinely ~\citep{Spoelstra2014,Smith2011}. In order to capture this aspect of charging behavior, agents are defined to have several areas in which they frequently charge. In these areas, agents regularly display the same type of activity. Each area contains one or more CPs, which the EV user had used in the past based on the data, forming a \textit{cluster} of the agent.

% center -> literally
% clusters -> set of CPs in a center

As such, the CP selection model consists of two steps; (1) a step that directs the agent to a cluster of frequent used CPs conditioned on time and space and (2) a second step that models the decision of a specific CP within the cluster. For each agent, its clusters and the behavior (captured in the dimensions of what, when and where) it exhibits in the cluster is purely extracted from its historical transactions. Each agent has one or more clusters, unique to that agent, and based on the historical transactions for each of these clusters three types of distributions are drawn: (1) arrival distributions, (2) connection duration distributions and (3) disconnection duration distributions. Given {\em where} and {\em when} an agent has previously charged, probabilities of where and when the agent will charge next are extracted from the data. This means no assumptions are made about either the centers or the behavior exhibited at their clusters.

The connection process takes the disconnection time and location of the previous connection as input and results in the time and cluster of CPs of the next connection. The CP selection process takes the cluster as input and results in the specific CP to connect to. The disconnection process takes the connection time and cluster as input and results in the time at disconnection.  A detailed description of the processes is given in section \ref{sec:submodels}.

Each cluster is uniquely attributed to a single agent and contains CPs which are geographically close together and at which the agent exhibits similar charging activity patterns. Even if two agents have clusters with the exact same CPs, the clusters are still attributed to each separate agent as a unique cluster.
Note, that in the simulation there is no explicit relation made between the type of geographic area (e.g. work or home location) and the activity patterns. This is also not needed, since the model captures behavioral patterns from the training data and therefore implicitly embeds area type information. For this reason the connection patterns of agents are not checked on expected patterns given the local area type.

\begin{figure}[ht!]
    \centering
    \includegraphics[width=0.6\textwidth]{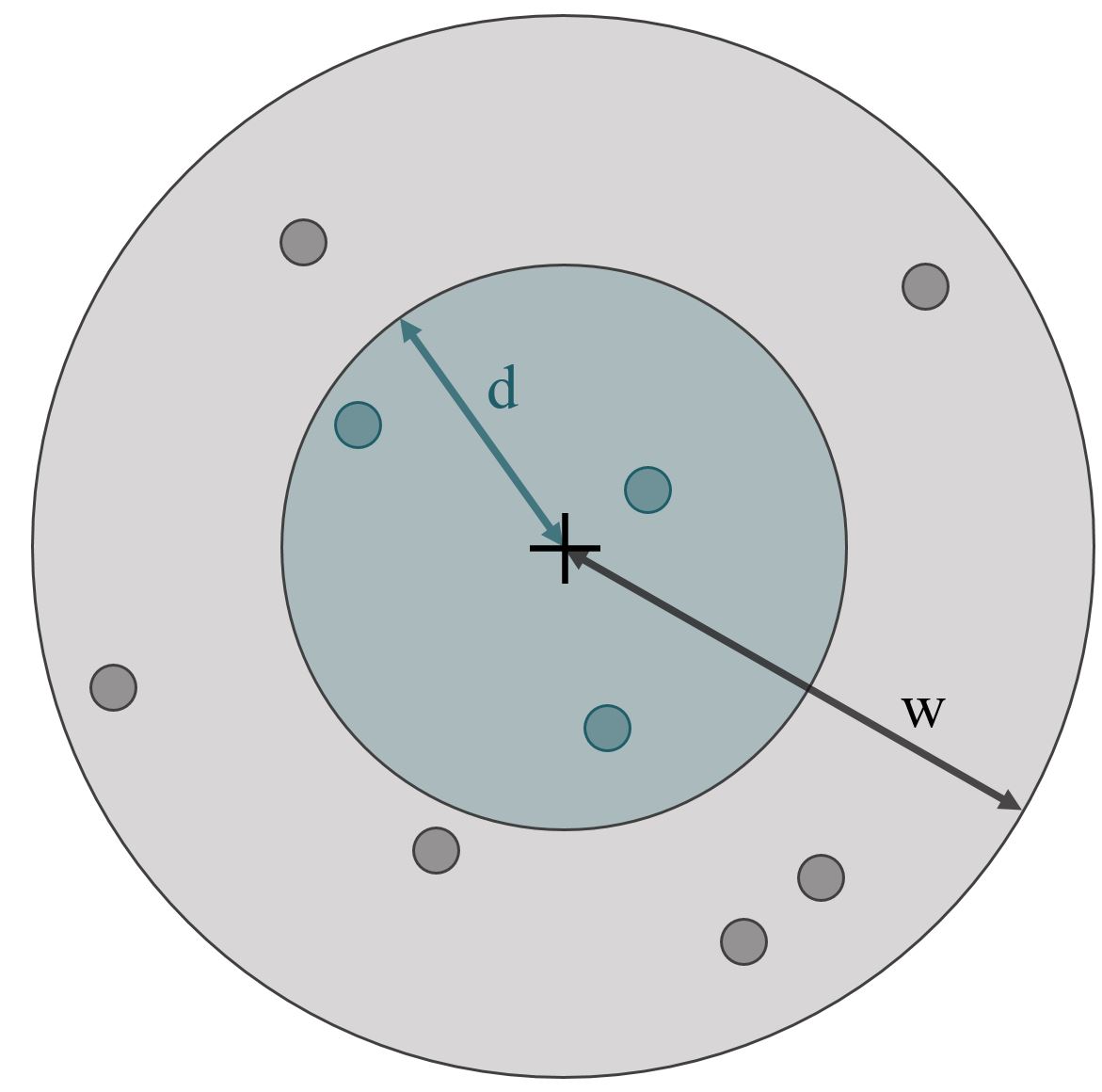}
    \caption{A cluster of an agent, where the cross indicates the center, the small blue circles indicate previously visited CPs and the small gray circles indicate previously unvisited CPs within range of the walking preparedness.}
    \label{fig:max_dist_walking_preparedness}
\end{figure}

The \textit{center} ($c$) is the average of the locations of the CPs ($p$) in the cluster, weighted by the number of charging transactions of the user at each CP in the data. This center is assumed to be an approximation of the location of the real destination of the user. The centers can, for instance, represent the EV user's home or work location, or any place where the user will charge their EV frequently. Implicitly this assumes that distance to the destination is the determining factor in choosing a CP. A schematic diagram of a cluster with its CPs can be seen in Fig. \ref{fig:max_dist_walking_preparedness}. Note that the center is located closest to the CP ($p_1$) with most transactions by the user.  The details on how to determine the clusters of an agent are discussed in \ref{sec:c2_creation_clusters}.

We formalize the calculation of the longitude ($c_{\text{lon}}$) and latitude ($c_{\text{lat}}$) of the center with the following equations:
\begin{align}
\label{eq:center1}
  c_{\text{lon}} &= \frac{\sum_{p \in c} p_{\text{lon}} \cdot p_{\text{tr} }}{c_\text{tr}} \\
\label{eq:center2}
    c_{\text{lat}} &= \frac{\sum_{p \in c} p_{\text{lat}} \cdot p_{\text{tr} }}{c_\text{tr}}.
\end{align}
Here $p_{\text{lon}}$ and $p_{\text{lat}}$ are the longitude and latitude of the CPs in the cluster. The number of transactions at a CP and within a cluster are indicated with $p_{tr}$ and $c_{tr}$, respectively.

%Fig. \ref{fig:max_dist_walking_preparedness} shows an example of a cluster of an agent.
In Fig. \ref{fig:max_dist_walking_preparedness}, the \textit{maximum distance} $d$ of an agent is defined as the maximum of the distances between the center and any of the CPs in the cluster plus 10\%. The reason for this surplus is to include optional CPs in case of sparse cluster formations.
 %reason

The \textit{walking preparedness} $w$ of an agent is defined as the maximum of $d$ of all clusters and based on sensitivity analysis the minimum walking preparedness $w_{min}$ ($w_{min}$ is set to 150 meters as can be seen in table \ref{tab:parameters agent}). The attribution of one  $d$ and $w$ per user, implicitly assumes a behavioral indifference to the time activity patterns. This means each agent has the same walking preparedness in the evening and during daytimes. Both $d$ and $w$ are kept constant during simulation.

\subsection{Design Concepts: Individual sensing and interactions}
\label{DC: Sensing}
During the simulation two driving factors are influencing the stochasticity of agent behavior in the simulation.
On one hand there is the exogenous factor of newly added CPs in the system, on the other hand there is the endogenous factor of competition of agents for CPs. The agents in the system have the ability to adapt their charging point selection to both factors by selecting newly added CPs or alternative CPs in case of occupancy. \\

Each agent is assumed to know the location of the CPs within the walking distance of its cluster. While the centers are fixed for an agent throughout a simulation, the size of a cluster in number of CPs of an agent may expand due to new CPs within the walking preparedness of an agent. As such, for each agent newly added CPs become part of the set of CPs of the cluster.

By definition new CPs are not part of the historical data of an agent and as such not part of the distributions from which the behavior is derived. In order to enable the agent to adapt to a changing environment, a habit probability ($p_h$) is added to the connection process that lets the agent select a CP either from its behavioral habits (the data) or to from the nearest available CP from its cluster center. This allows agents to adapt their behavior to use new, closer CPs. Connections to the new CPs do effect the behavioral distributions and as such the agent does learn from using new added CPs.

The knowledge of CP location and availability is designed such that it complies with the real-world system. EV drivers may access information resources that share CP availability data with a certain level of accuracy. As such, agents have full knowledge of CP locations while it is assumed that EV users have some limitations in knowledge about availability (either due to not using apps, inaccuracies in the apps, or competing EV drivers starting to charge just earlier than the agent). Next, the EV users are also not aware other users' attempts to connect to the CPs that it is currently connected to. As such, in the simulation model there is no information sharing between users that allows for collaboration between users to reduce competition in the system.

Given the these information limitations, we have chosen to model the awareness of CP occupation as follows: \emph{an agent is aware of the CPs locations, but not aware of the occupancy of a specific CP until it selects one to connect to}. If a selected CP is occupied, a record is stored of a failed connection attempt to be used for later analysis. Failed in this case implies failed to select to a preferred CP, while charging elsewhere might have worked.

The historical failed connection attempts do not influence future behavior, yet successful connection attempts are added to the distributions that form the behavioral properties. As such, the agents are able to learn and adapt to a changing environment. A more extended explanation of decision making can be found in \ref{sec:submodels}.

\subsection{Design Concepts: Observations}
\label{sec: Observations}
The set simulated charging transactions $Tr=\{tr_1...tr_n\}$ is the main output of the model and consists of a tuple $\{u_j,cp_i,t_{s},t_{e} \} $ in which $u_j$
 the agent, $cp_j$ is the Charging Point, $t_{s},t_{e}$ are the start and duration of the connection. These transactions can then be analyzed and transformed into other required outputs, such as charging infrastructure key performance indicators \cite{Helmus2018a}. Next, a set of failed transaction attempts is stored $F=\{f_1...f_n\}$ as well in a tuple of  $\{u_j,cp_j,time_{start}\}$. The failed connection attempts can be used to reveal dynamics in the system that can not be discovered in transaction data.

From the simulated charging transactions $Tr$, activity patterns are created using the charging transactions that summarize the behavior of agents. An activity pattern captures the activity of a (group of) agents or CPs over the 24 hours of a day in bins of $T$ minutes.
%It is constructed in the following way: We defined a $bin=\{T_x,T_y\}$ as an interval and $Bins=\{bin_1,..bin_k\}$ as a set of bins. A parameter is $\delta$ is defined as the binsize and  based on this a vector sized $k$ of $$bins={[00:00,00:00+\delta),[00:20:00:40),...[00:00-\delta,00:00)}$$ was created.
Based on simulation metrics, a default bin size of 20 minutes was chosen, see \ref{sec:base_model_evaluation}. In addition to the aforementioned observations, several other metrics are stored as well, see \ref{tab:metrics}.

% % % % % % % % % % % % % % % % % % % % % % %
%            INITIALIZATION                 %
% % % % % % % % % % % % % % % % % % % % % % %

\subsection{Details: Initialization}
\label{sec:initialization}
\begin{figure}[H]
      \includegraphics[width=\textwidth]{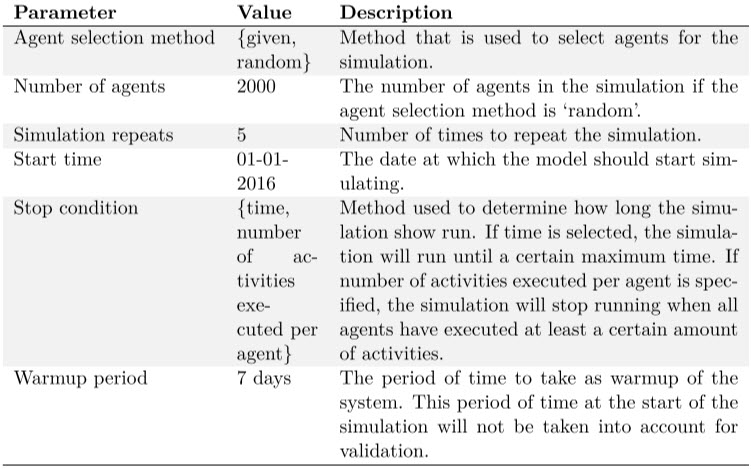}
      \caption{Parameters present in the model concerning the simulation}
		\label{tab:parameters simulation}
  \end{figure}

%This section will describe the process of initializing the agents and the environment.
The simulation handler receives the input parameters as specified in tables \ref{tab:parameters simulation}, ~\ref{tab:parameters data},~ ~\ref{tab:parameters agent} and  ~\ref{tab:parameters clustering}. While these tables contain default values (motivated in section \ref{sec:base_model_evaluation}), for each of the input parameters, the values can be changed for a simulation run. Unless explicitly stated, experiments were run with default values.

Using the aforementioned parameters, the raw data is loaded and pre-processed. In the preprocessing invalid entries are removed from the dataset in line with \cite{VanDenHoed2014}. After preprocessing, information about a CP; the GPS location, the number of sockets per CP, the parking zone of a CP is gathered from the data. Next, an instance of the environment is created, where all CPs present in the dataset are loaded and set to be unoccupied. The set of transaction is split into a training and test set based on a split data that successively defines start and end dates for both training and test in the input parameters.
Next, instances of the agents are created according to the agent selection method specified in the parameters (see agent selection method in table \ref{tab:parameters simulation}). At initialization, an agent first gets a unique ID after which is behavior is captured, summarized and stored. This is done by generating the clusters and distributions of the agent using the charging transactions of the agent.

\subsubsection{Summarizing Behavior: the Creation of the Distributions}
\label{sec:distributions}
An important step during initialization is the creation of the behavioral distributions of each individual agent.
As mentioned in Section \ref{sec:process overview} every agent in the simulation has three types of distributions:
\begin{itemize}
    \item An arrival time distribution per cluster;
    \item A connection duration distribution per cluster; %(time interval with length of the bin size);
    \item A disconnection duration distribution per arrival time bin.
\end{itemize}

The precision of the simulation model, and thus the number of discrete intervals in each distribution, is determined by the bin size $T$ (see table \ref{tab:parameters data}).

Each cluster of an agent has an arrival time distribution. This distribution is used to sample a start time of connection during simulation and at the initialization of the simulation (see Section \ref{sec:initial state} on how this is used). This distribution ranges over 24 hours of a day and the bins indicate a time interval within a day (for example between 1:00pm and 1:20pm). The value of that bin indicates the number of occurrences with which an agent connected to any CP in a specific cluster in the data. Next, for each cluster of an agent a connection duration distributions is setup from the duration of transactions within the specific cluster. This distribution is used to sample the connection duration.

While the duration distribution being time based has a range that depends on the maximum duration of connection in the data and thus may exceed 24 hours. This implies that for each agent there are as many arrival time and connection duration distributions as clusters.

%The disconnection duration $dd$ is calculated from the difference $dd_{n,n+1} = \Delta \big( \text{time}_{\text{start,n+1}}-\text{time}_{\text{start},n}-\text{time}_{\text{duration},n} \big) $ between two subsequent transactions $Tr_{n}, Tr_{n+1}$ regardless of the clusters. % TODO find out >>>> regardless or conditioned on the clusters??
The disconnection duration distributions are conditioned on the start time of the first transaction $Tr_{n}$, such that for each bin of start times in the set of transaction of an agent a disconnection distribution is developed.

When the agent disconnects from a CP, a sample is drawn from the relevant disconnection duration distribution. This is used to
determine the time at which the agent will start its next connection.

\subsubsection*{Initial State}
\label{sec:initial state}
Once the behavioral distributions are created, the initial state of the agents can be set. This state contains the following variables: is connected, time next activity, active center and active CP (see figure \ref{tab:entities}).

First, for each agent it is determined whether it is connected at the beginning of the simulation. To do so, for each agent the overall activity pattern based on the time bins of a day is calculated and normalized for the total number of sessions. This results in a probability distribution of being connected. From that the value of the bin in which the start time of the simulation is taken as probability that the agent is connected at the start of the simulation. What remains is to determine when the agent will (dis)connect and where the agent is connected (if it is connected) or will connect to (if it is disconnected).
\bigbreak
If an agent starts out disconnected, a time and place for its first connection must be set. To do so, a cluster with probabilities according the number of transactions in the data for each cluster is chosen. Once this cluster is sampled, an arrival time from that cluster's arrival distribution is sampled as well.  As a result, the time of the next activity of the agent, namely connecting to a (yet to be determined) CP within the sampled cluster.

If an agent starts out connected it is determined to which CP it is connected to as well as when its next activity of disconnecting will take place. For each center the number of occurrences that the agent was connected at a CP within this cluster is calculated given the time of day of the simulation start. This results in a probability distribution for the user conditioned on this specific time.  From this distribution a specific cluster is sampled. The next step is to sample an arrival time from the cluster's arrival distribution. The sampled arrival time is set in the 24 hours previous to the start of the simulation. A CP is then selected using the selection process as described in Section \ref{sec:process overview} (and in more detail in Section \ref{sec:submodels}). Finally, from the cluster's connection duration distribution the duration is sampled to determine when the agent will disconnect from this CP and set this as the time of the next activity.

\subsection{Details: Input Data}
\label{sec:input data}
The dataset used for this data-driven model is gathered by the IDOLAAD project \cite{UniversityofappliedsciencesAmsterdam2016} and contains over 5.6 million charging transactions conducted by approximately 40 thousand EV users at around 5 thousand CPs located throughout metropolitan areas of the G4 cities in the Netherlands. Additional information about (the creation and cleaning of) the dataset can be found in~\cite{VanDenHoed2014}. The size of the dataset as well as the geographical diversity provides a reliable base for understanding the factors influencing the charging behavior of EV users on public charging infrastructure in the Netherlands.

% % % % % % % % % % % % % % % % % % % % % % %
%           SUBMODELS SECTION               %
% % % % % % % % % % % % % % % % % % % % % % %

\subsection{Details: Submodels}
\label{sec:submodels}
A high level of modularity was achieved by modeling each process (connection, disconnection and CP selection) with its own internal working, independent of the rest of the simulation. Therefore, each individual process can be adjusted, updated and improved without interference from the rest of the processes. Also new processes, like State of Charge (SOC) modeling that allows for transaction volume simulation, may be added if desirable.

This section provides a detailed description of three processes: the connection process, the CP selection process and the disconnection process. Next it provides a detailed description of how activity clusters of agents are created. Each process has two functions; first is sets the state of the agent and second it generates a timestamp for the next event of the agent in the entry of the queue of the simulation handler. An overview of the input, output and system interaction of each of the three processes is shown in Table \ref{tab:process_overview}.

\begin{table}[H]
    \centering
    \begin{tabularx}{\textwidth}{p{2.2cm} p{3cm} p{3cm} X }
        \textbf{Process} & \textbf{Input} & \textbf{Output} & \textbf{System interaction} \\\hline
    \rowcolor{Gray}
    \hline
        Connection process & time of connection and cluster of previous connection & cluster of next connection & - \\
        CP selection process &  Cluster and time of connection &  CP of connection and time of disconnection & Adds agent to CP at time of connection. \\
     \rowcolor{Gray}
        Disconnection process & Time at disconnection & disconnection duration & Removes agent from CP at time of disconnection. \\
    \end{tabularx}
    \caption{An overview of the input, output and system interactions of each process.}
    \label{tab:process_overview}
\end{table}

The \emph{connection} process determines at which cluster an agent will connect based on the given connection which is the result of the disconnection process. Once the next cluster has been determined, it feeds forward its output to the CP selection process. The \emph{CP selection} process determines at which CP in the cluster the agent will connect. When a CP has successfully been selected (indicated with the green arrow in Figure \ref{fig:agent execute activity}), the agent becomes connected and the next process to call upon is the disconnection process. When the CP selection process fails (the red arrow in Figure \ref{fig:agent execute activity}) the agent remains disconnected and the connection process is executed upon again after $t_r$ (on default 20 minutes) time (see table \ref{tab:parameters agent}). The \emph{disconnection} process is called at the disconnection time when an agent disconnects from its current CP and feeds forward to the connection process. To disconnect the agent, the socket of the CP to which the agent was connected will be freed.

\subsubsection*{Connection Process}
\label{sec:connection process}
The connection process simulates at which cluster the next connection will be given the time that the agent wants to connect. This connection time is calculated from the disconnection duration as a result of the disconnection process. The agent calls upon this process from a disconnected state. \\
Conditioned on the time of connection, each of the clusters of the agent is given a probability according to the frequency with which it arrived at the cluster at that time. Thereafter, a cluster is drawn according to those probabilities and this will be the cluster the agent will connect to next. The clusters are user specific and that in the assignment of a user to a cluster. The connection process does not check on availability, as such this process does not fail or iterate.

\subsubsection*{CP Selection Process}
\label{sec:selection process}
The CP selection process determines which CP an agent should connect to when the cluster that it aims to connect to is given as a result of the connection process. It is executed after the connection process. In this version of the model there are two possible selection methods: habit based or distance based as described in section \ref{DC: Sensing}. \\

\begin{figure}[H]
      \includegraphics[width=\textwidth]{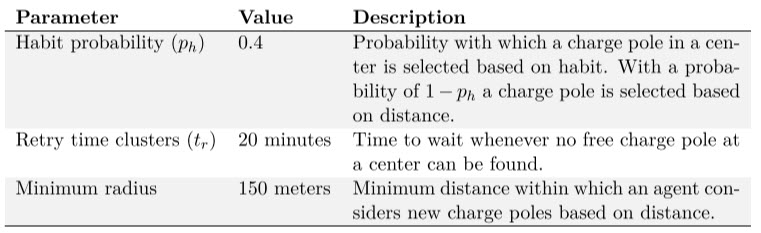}
      \caption{Parameters present in the model concerning the agents behavior}
		\label{tab:parameters agent}
  \end{figure}

With a chance determined by the habit probability parameter $p_h$, with default value $0.4$ (see table \ref{tab:parameters agent}), the agent will select a CP based on habit.
If the agent selects a CP based on habits, a CP is selected with a chance directly proportional to the number of transactions the agent had at that CP in both the training data and in the simulation up to that point. As such, agents are able to learn and adapt behavior to new CPs in the environment. With a chance of $1 - p_h$ the next CP to select is selected based on distance. That is, the CP closest to the agent's center (while still within the range of the agent's walking preparedness) with an available socket is selected. The distance based selection method thus assumes that the agent has an awareness of availability of the CPs at that time.

It is possible that the agent chooses an occupied CP using the habit method. In this case a tuple if added to the set of failed connection attempts $F$ (see also \ref{sec: Observations} ) storing the relevant information of the failed connection attempt. The CP selection process is iterated using the habit method, until either successful CP selection or failure of selection of any of the habitual CPs. In the latter case, the CP selection process based on distance is used.

The final result of the CP selection process has two possible options; either the agent has selected a CP successfully or it failed to select a CP. If the agent was successful its state is changed to connected and a socket of the CP to which it connects becomes occupied. The socket to connect to within a CP is simply the first one available, since the sockets of a single CP are interchangeable it does not matter which one is picked.
In order to find the disconnection time, a sample is taken from the connection duration distribution that belongs to cluster the agent is connected to. The connection duration distribution is conditioned on the bin of the current time in order to get a duration for the current connection (and thus decide when to disconnect). The time for the next activity of the agent will then be the current time plus the sampled duration.

A check on the plausibility of the connection duration is done to avoid skewed behavior. The connection duration is considered plausible if within the bin of the resulting disconnection time, the agent is known to have disconnected from the cluster in the training data. If this is not the case, the simulation checks the bins preceding and succeeding the bin originally sampled. If the newly chosen bin is still invalid, a check the second bins preceding and succeeding the originally chosen bin is performed. This process continues until a valid bin is found. Note that the minimal connection duration is at least one bin size long to avoid succeeding repeats of the same behavior.

If the agent was unsuccessful this means that all relevant CPs are at the moment occupied and the connection process will be called upon again after the retry time $t_r$ (20 minutes default) (see table \ref{tab:parameters agent}). This will continue until an available CP is found.

\subsubsection*{Disconnection Process}
The disconnection process changes the state of the agent to disconnected and updates the environment such that the socket of the CP at which the agent was connected becomes free again. It  determines the disconnection duration and with that the time of the subsequent connection. The disconnection process takes the time of disconnection as input. A sample is taken the disconnection duration distribution conditioned on the time bin of disconnection time. This results the duration of the disconnection and combined with the current time this process determines the time at which the agent start its subsequent connection. However, check is needed on the upcoming connection time whether this time is valid for that agent. \\
For example, if an agent is known to connect during work hours at a certain location, it is not plausible that the agent connects there at 2am. We consider a time to be valid if within the bin of that time, the agent is known to have connected from the cluster in the training data. If this is not the case, the simulation checks the bins preceding and succeeding the bin originally sampled. If the newly chosen bin is still invalid, a check the second bins preceding and succeeding the originally chosen bin is performed. This process continues until a valid bin is found. Once a valid bin is selected, the time of the next activity (connecting to the system) of the agent is set to be the current time plus the found disconnection duration. An extra restriction is made that the connection duration should be at least one bin size long (see \ref{tab:parameters data}) to avoid repeated short connection and disconnection times.

\subsubsection{Cluster formation}
\label{sec:c2_creation_clusters}
\begin{figure}[H]
      \includegraphics[width=\textwidth]{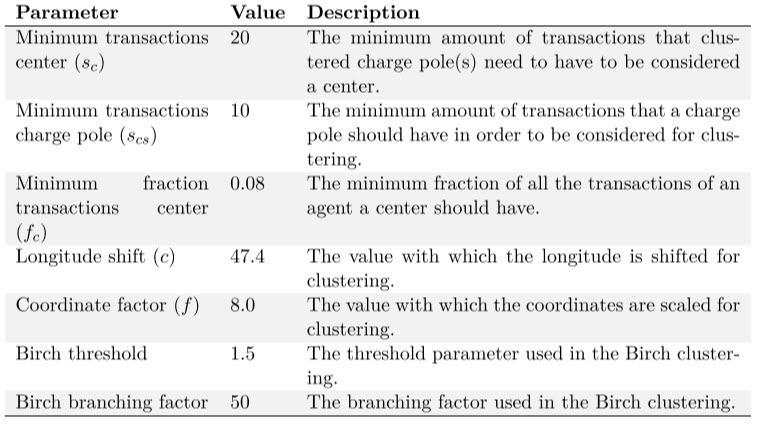}
      \caption{Parameters present in the model concerning the clustering agents charge poles}
		\label{tab:parameters clustering}
  \end{figure}

% center -> literally
% clusters -> set of CPs in a center
Based on the transaction data, for each individual agent the CPs that this agent uses are clustered into one or more clusters. Clusters are formed with CPs that are close together (physical location) and where the agent exhibits similar behavior (activity patterns). The clustering requires several parameters, which are all listed in table \ref{tab:parameters clustering}.

%each cluster has a center based on the weighted lon lat of all sessions

In the cluster formation all CPs which the agent has visited more than $s_{cs}$ (on default 10) times in the training data are considered. For each of those CPs the activity pattern of the agent at this CP is determined as well as the longitude and latitude. This data is the input for the clustering algorithm. However, some pre-processing is needed to ensure that sensible clusters are created. First, the longitude and latitude are scaled to same range of values. The longitude is shifted with $h$ (see \ref{tab:parameters clustering}), which is the mean of the latitude values of all CPs in the data minus the mean of the longitude values.

Second, the influence of distance and time need to be equalized in the clustering input, do the the fact that the distance has only two variables (longitude and latitude) whereas the number of time variables is equal related to the bin size of the activity pattern.
Intuitively CPs will now be clustered when they have similar activity patterns, longitude and latitude. This means that CPs which are physically close to each other and where the user exhibits the similar temporal behavior will be clustered together.
However, to increase the influence of distance (rather than the temporal behavior) on the clustering we choose to multiply the longitude and latitude with a scaling factor $f$ (see \ref{tab:parameters clustering}). The resulting data is clustered using the BIRCH algorithm without a pre-specified number of clusters~\citep{Zhang1996}. The input parameters for the algorithm were tuned to the values found in \ref{tab:parameters clustering}. A sensitivity analysis and argumentation for these values is given in Section \ref{sec:base_model_evaluation}.

The purpose of the SEVA model is to reveal the dynamics of the habitual EV charging behavior, rather than incidental behavior \cite{Helmus2018a}, Therefore, incidental sessions (e.g. near points of interest) should not be taken into account during formation of habitual behavior. To filter out these sessions, an agent needs to have a minimum number of transactions $S_{cs}$ at a specific CP for that CP to be considered for clustering. The agent must also conduct at least a fraction $f_{c}$ (default $0.08$) of its total charge sessions at that CP. Groups of CPs which satisfy these criteria form the clusters of the agent. Agents are only valid, and thus present in the simulation if they have at least one cluster. A center itself is located at the mean location (longitude and latitude) of all CPs within that cluster, weighted by the number of transactions at each of the CPs as explained in Section \ref{sec:entities}.

% % % % % % % % % % % % % % % % % % % % % % %
%            METRICS SECTION                %
% % % % % % % % % % % % % % % % % % % % % % %

\clearpage
\section{Model Metrics}
\label{sec:metrics}
In this section the metrics that capture the output of the model are described, see also table \ref{tab:metrics}. The metrics are divided into clustering, run time and validation metrics.

%do we need to use subsections?
\subsection{Clustering Metrics}
The method of clustering the CPs of the agents into clusters is fundamental for the simulation model, as it captures where agents exhibit which behavior. Therefore, to be able to measure and verify the outcomes of the clustering process, several clustering metrics are generated.

For each agent the we store:
\begin{itemize}
  \item \emph{number of CPs in a cluster}
  \item \emph{number of clusters}
\end{itemize}

Furthermore, we store the total number of CPs in the simulation used by at least one agent. Lastly we take into consideration the physical size (radius) of the clusters (see table \ref{tab:metrics}), as explained in section \ref{sec:entities}.

\subsection{Run Time Metrics}
We gain insight about the computation time of the simulation by calculating  run time metrics. As the model is developed for the purpose of doing large-scale simulations, it is important that computation time stays within bounds even when the size of the experiments grows. For this purpose two metrics are introduced that collect the time of the most computationally intensive parts of the model, namely the initialization time and simulation execution time.

\subsection{Validation Metrics}
\label{subsect:validation}
Finally, metrics are stored to be used for validation of the simulation results. To do so, first the data is divided into a \emph{training} set and a \emph{test} set based on a longitudinal split on a specific date, see table \ref{tab:parameters data}. In this paper, data from  2014-1-1 to 2016-1-1 is used as training data to initialize agent behavior and 2016-1-1 to 2017-1-1 is the test data. Note, that as the simulation is stochastic, validation can be performed on both the training as well as the test dataset. We also consider two forms of validation (1) validation of CP activity and (2) validation of agent activity.

To determine whether the simulated data matches the training and test data, a validation metric is needed. We use the activity patterns of the real data (test and training) and compare these against the activity generated by the simulation. To compare activity patterns, we calculate the error between every bin in the activity pattern of the simulation result against the corresponding bin in the activity pattern of the real data. The maximum possible error for bin $i$ of the simulated data $b^s_i$ and bin $i$ of the real data $b^r_i$ is:

\[ d_{max} = max(b^s_i, 1- b^s_i, b^r_i, 1-b^r_i) \]

We calculate the error between bin $i$ of the simulated data $b^s_i$ and bin $i$ of the real data $b^r_i$ as:

\[d_i =  \frac{| b^s_i - b^r_i |}{d_{max}}\]

The error value of the whole activity pattern can then be viewed as the average of the error values of all its bins. An agent then has an average error per cluster - calculated by comparing cluster activity in simulated and real data. For an experiment with multiple runs we then calculate the mean error of cluster activity for each agent.

For the CP validation, we calculate the same form of error based on the activity patterns at that CP. However, we need to be careful here, since the data contains transactions from users that are not simulated. Therefore, only CPs used by simulated agents are considered, and only transactions made by simulated agents are considered.

Clearly lower error values indicate better validation. However, at the population scale it can be difficult to interpret the meaning of the error. For the reader we provide example validation values with corresponding activity patterns in appendix \ref{Appendix validation} figure \ref{fig:activity pattern examples}.

%%%% ---------------------------- %%%%
%%%% ---------------------------- %%%%
%%%% Model Evaluation             %%%%
%%%% ---------------------------- %%%%
%%%% ---------------------------- %%%%
\clearpage
\section{Sensitivity Analysis and Model Evaluation}
\label{sec:base_model_evaluation}
\begin{figure}[H]
      \includegraphics[width=\textwidth]{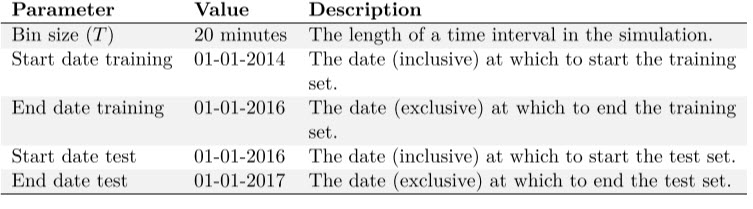}
      \caption{Parameters present in the model concerning data pre-processing}
      	\label{tab:parameters data}
  \end{figure}

In this section various experiments regarding model sensitivity analysis and model evaluation are discussed. We use One Factor at a Time (OFAT) sensitivity analysis and the initial default values are stated in \ref{tab:parameters simulation},~\ref{tab:parameters data},~\ref{tab:parameters agent} and~\ref{tab:parameters clustering}. We test the sensitivity of a number of system parameters by checking their effect on the validation metric. Note, we conducted analysis of all the simulation parameters but in the main body of the paper we include only the more interesting and relevant results. For other parameters please refer to section~\ref{Appendix validation}. 

% The first step is to determine the optimal values for the simulation setup in number of agents and simulation repeats and the binsize used in the simulation. Subsequent analyses were performed given these default values. First, the clustering parameters were examined since the clustering contains an essential part of the simulation model. Finally, the parameter sensitivity of the CP selection process was examined as described in \ref{sec:selection process}.

For each of the experiments we consider four different validation metrics.
\begin{enumerate}
\item \emph{CP training} represents the mean error over all CPs for simulated data vs training data.
\item \emph{EV training} represents the mean error over all agents for simulated data vs training data.
\item \emph{CP test} represents the mean error metric over all CPs for simulated data vs test data.
\item \emph{Ev test} represents the mean error metric over all for simulated data vs test data.
\end{enumerate}

\subsection{General Parameters}
%bin size
\label{subsec: simulationsetup}
We tested three general parameters of the simulation, {\em number of users}, {\em number of simulation repetitions}, and the {\em the activity distribution bin size}. The first two parameters are included in the appendix %(see \ref{sec:v Appendix validation}) 
and the final one is included in the main body of the paper.

\begin{figure}[H]
    \centering
    \begin{subfigure}[b]{\textwidth}
        \includegraphics[width=\textwidth]{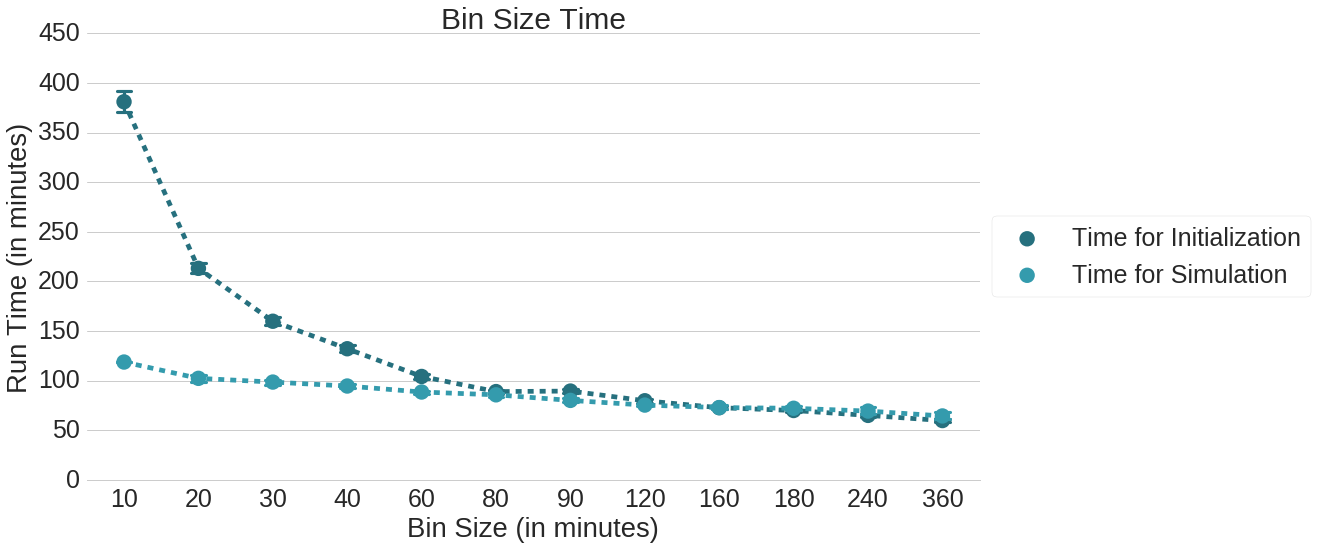}
        \caption{}
        \label{fig:validation bin size time}
    \end{subfigure}
    \begin{subfigure}[b]{\textwidth}
        \includegraphics[width=\textwidth]{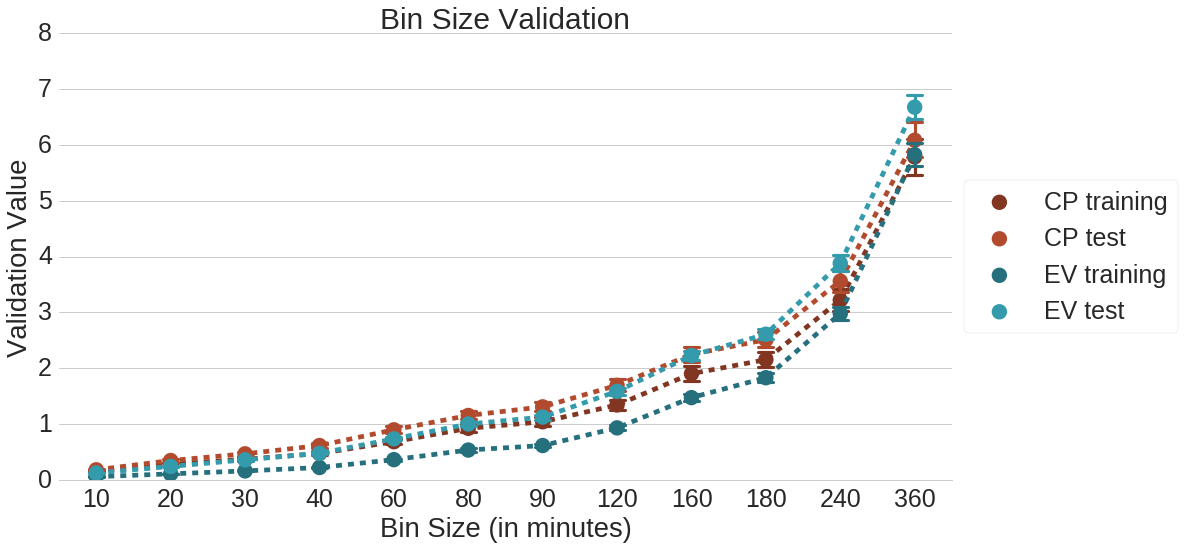}
        \caption{}
        \label{fig:validation bin size validation}
    \end{subfigure}
    \caption{The results of the experiment that varies the bin size. Note that the horizontal axes do not have a linear scale. Per value of the parameter we plot \textsc{(a)} the mean run time and \textsc{(b)} mean validation value with the 95\% confidence interval.}
    \label{fig:validation bin size}
\end{figure}

Varying the bin size affects both simulation run time as well as the validation metric, since it determines how fine-grained the distributions of behavior are. The decision for a default bin size is a trade-off between validation requirements and run time. The results of bin size variation, both simulation run time and the validation value, can be found in Figure~\ref{fig:validation bin size}. From Figure~\ref{fig:validation bin size validation}, it can be seen that the time for initialization decreases exponentially and linearly for simulation. Next, it can be seen that the model validation metrics increase exponentially as bin size increase. Based on these numbers, a default bin size of 20 minutes was chosen as this reduces the run time by half compared to a bin size of 10 minutes, while the validation values only increase slightly.

\subsection{Clustering and Cluster Analysis}
In this section we tested the parameters associated with the clustering. We tested : (1) the coordinate factor, (2) the minimum number of transactions per CP, (3) BIRCH threshold and the (4) BIRCH branching factor. The main body includes the first three of these and the last is included in the appendix.

The clustering process is an important concept of the simulation design and therefore (1) parameters on data generation for the clustering and (2) the parameters of the clustering need to be taken into account in the sensitivity analysis. This analysis focuses on the effect of parameters on behavioral measures, since the parameters affect the generation of the agents' behavioral distributions. In these experiments the average number of CPs in a cluster, the number of clusters per agent, the average maximum distance and the average walking preparedness are analyzed as a result of parameter changes. We look to achieve an average walking preparedness of 250m for the clusters. This means the average agent will be willing to walk for 4 minutes from their parking spot to their activity center (e.g., home).

Recall that the clusters are defined as the set of CPs and the centers as the weighted average (longitude and latitude) of all the sessions within the cluster (see \ref{IndividualDecisionMaking}). There is therefore a one to one correspondence with the number of centers and the number of clusters.
% center -> literally
% clusters -> set of CPs in a center

\begin{figure}[H]
    \centering
    \includegraphics[width=\textwidth]{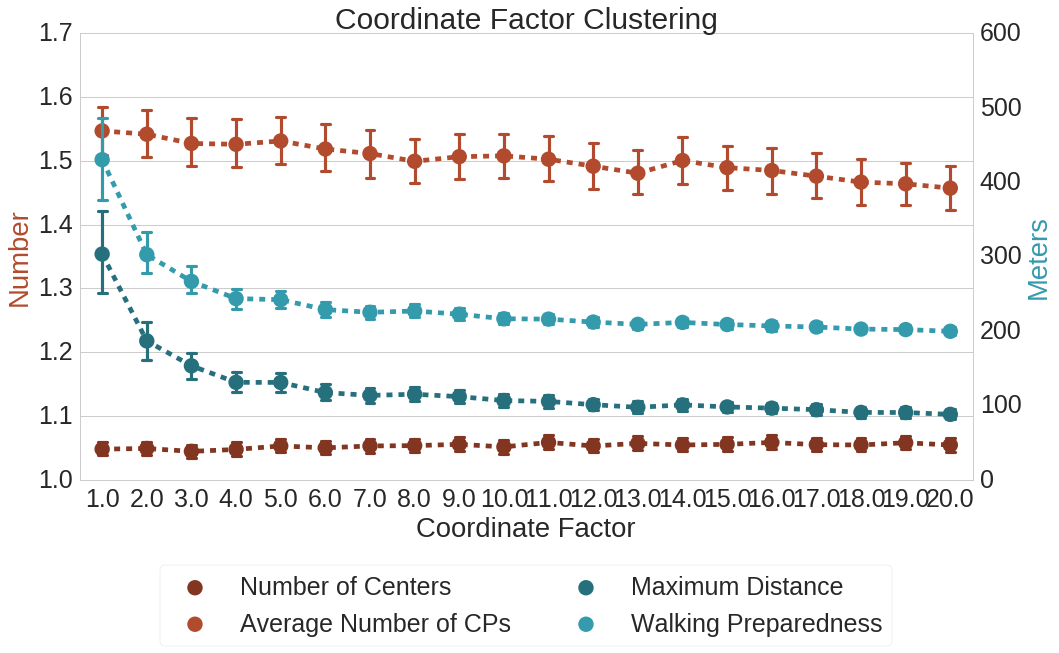}
    \caption{The influence of the scaling factor parameter. Per value of the parameter we plot the mean values with the 95\% confidence interval.}
    \label{fig:scale lon lat}
\end{figure}

Figure \ref{fig:scale lon lat} shows the sensitivity of the coordinate factor (see \ref{sec:c2_creation_clusters}). We can see that at the lower range of values there is a strong influence on the maximum distance and walking preparedness of the agents. The number of centers and number of cluster is relatively insensitive to changes. We have picked the value for this parameter to be $8$, since at this value the changes have leveled off.

%%%%%%%%%%%%%%%%%%%%%%%%%%%%%%%%%%
%   MIN TRANSACTIONS CSS
%%%%%%%%%%%%%%%%%%%%%%%%%%%%%%%%%%

\begin{figure}[ht!]
    \centering
    \includegraphics[width=\textwidth]{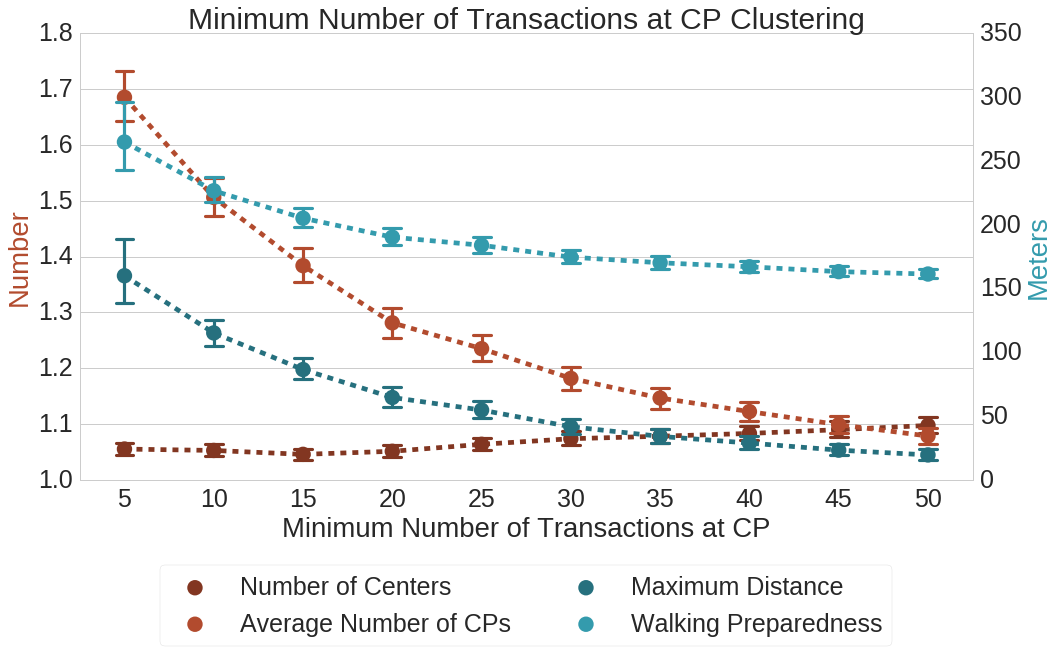}
    \caption{The influence of the minimum number of transactions per CP. Per value of the parameter we plot the mean values with the 95\% confidence interval.}
    \label{fig:minimum transactions css}
\end{figure}

 It is expected that the number of CPs per cluster and thereby walking preparedness and maximum distance all decrease as $S_{cs}$ (see section) increases. In Figure \ref{fig:minimum transactions css} the result of the sensitivity analysis is shown. It can be seen that this parameter has a strong influence on the number of CPs per cluster and the maximum distance, while having less influence on the walking preparedness and the number of clusters per user. We picked the value $10$ based on these results, because we do not want the number of CPs per cluster and the maximum distance to drop too far.

%%%%%%%%%%%%%%%%%%%%%%%%%%%%%%%%%%
%   BIRCH THRESHOLD
%%%%%%%%%%%%%%%%%%%%%%%%%%%%%%%%%%

\begin{figure}[ht!]
    \centering
    \includegraphics[width=\textwidth]{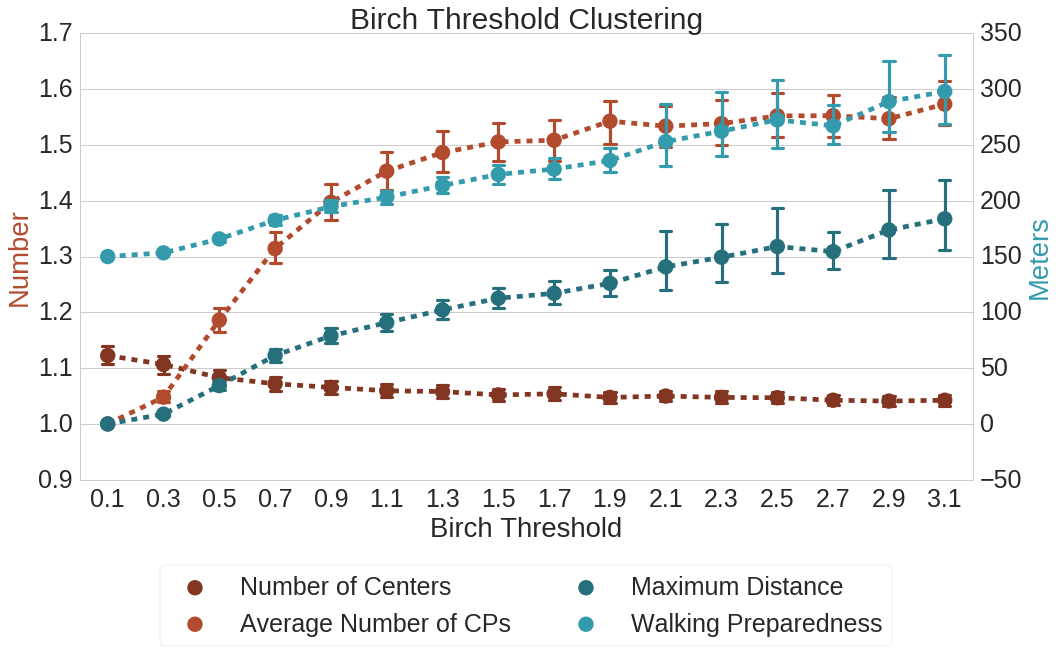}
    \caption{The influence of the BIRCH clustering parameter on various metrics. Per value of the parameter we plot the mean values with the 95\% confidence interval.}
    \label{fig:BIRCH threshold}
\end{figure}

Now that the parameters for cluster input data are set, the parameters for the BIRCH clustering algorithm can be analyzed. The BIRCH threshold is an input parameter of the BIRCH clustering algorithm ~\citep{Zhang1996} that sets requirements to form a sub cluster in the clustering tree. As such, increasing the threshold increases the cluster size in terms of the number of CPs and distance. The influence of varying this parameter on the clustering is shown in Figure \ref{fig:BIRCH threshold}. As expected, an increase of the threshold affects the average number of CPs per cluster and the maximum distance and walking preparedness. The error bars of all metrics except for number of clusters tend to increase as the threshold increases. This indicates increasing variance of cluster properties in the user population. A value of between $1.5$ and $1.7$ provides stable dynamics (low variance, levelling off) and also matches walking preparedness of 250m. In our simulations we chose to use a value of 1.5.

\subsection{Selection Process}
The selection process is also a fundamental element of the simulation model and thus its sensitivity is analyzed. Three parameters can be adjusted in the selection process (see also Section \ref{sec:selection process}): (1) minimum radius $d$ of the CPs taken into consideration for CP selection (2) the habit probability and (3) the retry time in case of no available CP at time of selection. The main body of the paper includes the second of these parameters, while the first and third are included in the appendix.

%%%%%%%%%%%%%%%%%%%%%%%%%%%%%%%%%%
%   HABIT PROBABILITY
%%%%%%%%%%%%%%%%%%%%%%%%%%%%%%%%%%
\begin{figure}[ht!]
    \centering
    \includegraphics[width=\textwidth]{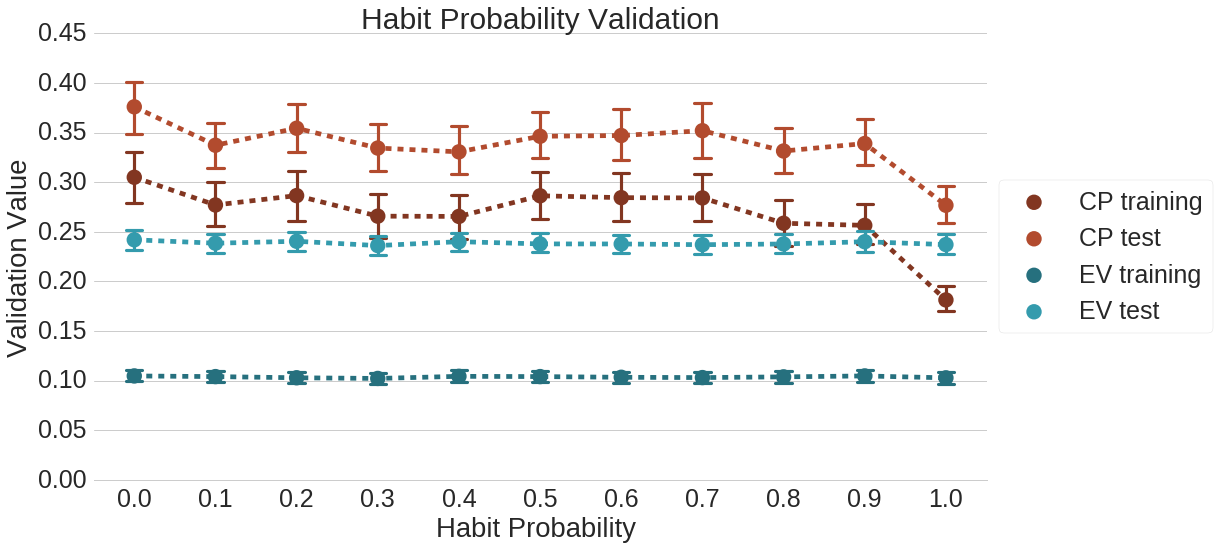}
    \caption{The influence of the habit probability. Per value of the parameter we plot the mean values with the 95\% confidence interval.}
    \label{fig:validation habit probability}
\end{figure}
The habit probability influences the probability of choosing a CP based on historical behavior as introduced in Section \ref{sec:selection process}. In Figure \ref{fig:validation habit probability} it can be seen that there is a clear dip in the CP validation at a habit probability of $1.0$. This can be explained by reasoning that with a habit probability of one, an agent acts strictly according to the data and as such we expect that validation improves. However at this value new CPs will never be used by the agents. In general, when the habit probability parameter is lower, new CPs within walking distance of an agent's cluster will be used more often. To allow for agents to learn to use new CPs, a value lower than one is required for this parameter.
Based on the low CP training and validation value and the agent learning ability, a value of of $0.4$ was chosen for the habit probability. With this value an agent with a cluster containing would be able to shift its preference to a more optimally placed CP by picking that one based on distance roughly 60\% of the time.

\subsection{Model Validation}
\begin{figure}[ht!]
  \centering
  \begin{subfigure}[b]{0.45\textwidth}
      \includegraphics[width=\textwidth]{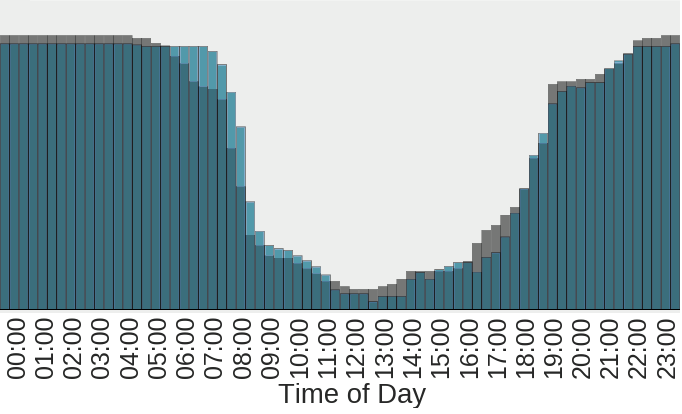}
      \caption{}
      % \label{fig:}
  \end{subfigure}
  ~
  \begin{subfigure}[b]{0.45\textwidth}
      \includegraphics[width=\textwidth]{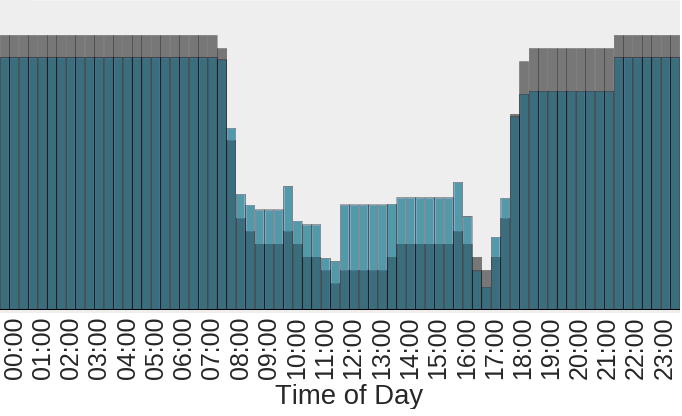}
      \caption{}
      % \label{fig:}
  \end{subfigure}

  \begin{subfigure}[b]{0.45\textwidth}
      \includegraphics[width=\textwidth]{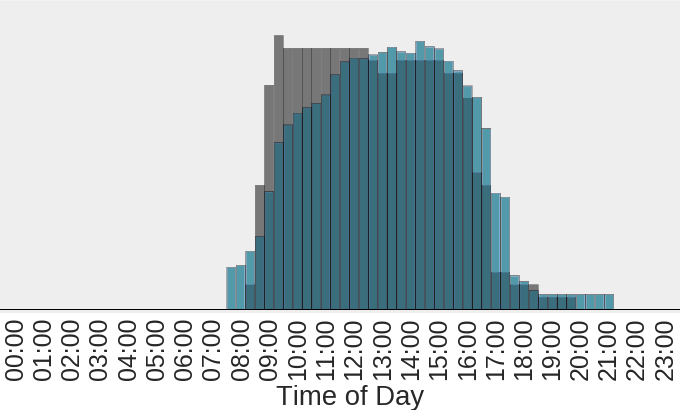}
      \caption{}
      % \label{fig:}
  \end{subfigure}
  ~
  \begin{subfigure}[b]{0.45\textwidth}
      \includegraphics[width=\textwidth]{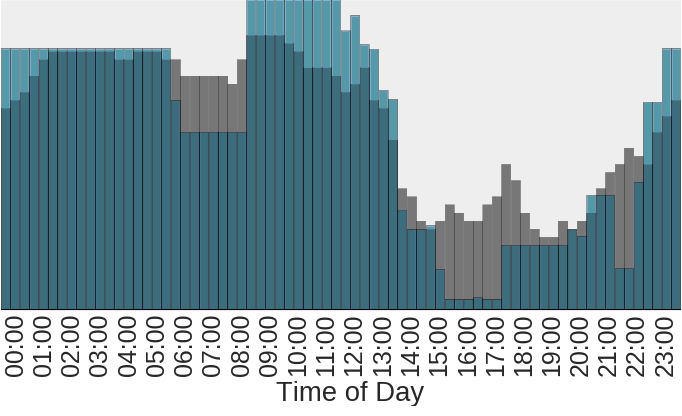}
      \caption{}
      % \label{fig:}
  \end{subfigure}
  \caption{Examples of activity patterns for the validation values \textsc{(a)} $0.102$, \textsc{(b)} $0.239$, \textsc{(c)} $0.296$ and \textsc{(d)} $0.333$. The real activity pattern is indicated in gray while the simulated pattern is blue. }
 \label{fig:activity pattern examples}
\end{figure}

% %%%%%%%%%%%%%%%%%%%%%%%%%%%%%%%%%%
% %   CORRELATION FIGURES
% %%%%%%%%%%%%%%%%%%%%%%%%%%%%%%%%%%
Having found the set of default parameters for the model, the overall model validation metrics can be run to check on validation. Several runs were performed using the chosen default values for the parameters and the following on average validation values were found: agent training validation value of $0.10$, agent test validation value of $0.23$, CP training validation value of $0.29$ and CP test validation value of $0.33$. Illustrations of the validation metric show that a value around $0.1$ shows a good match of the activity pattern (see Figure \ref{fig:activity pattern examples} in Section \ref{Appendix validation}). The CP training validation values are higher, due to the fact that CP activity patterns are dependent on the decision making of the agents and are therefore more sensitive to the interaction of the agents. 

\begin{figure}[ht!]
    \centering
    \includegraphics[width=\textwidth]{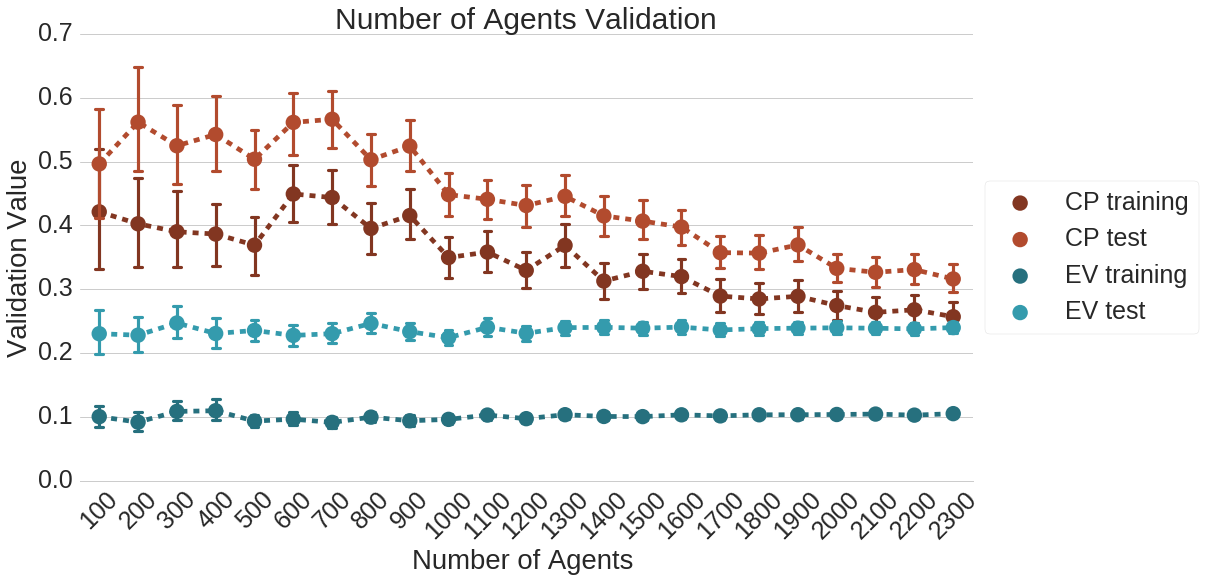}
    \caption{The results of the experiment where the number of agents is varied. Per value of the parameter the mean validation values with the 95\% confidence interval is plotted.}
    \label{fig:validation number of agents}
\end{figure}

To further validate the model we examine the effect of an increasing number of users in the system on the validation metric . Experiments were performed with gradually increasing number of users from 100 to 2300 in steps of 100. The effect of varying the number of agents in the system on the validation values can be seen in \ref{fig:validation number of agents}. The results show that the competition that occurs within the system, when more agents are present, does not influence the overall validation.

As the number of agents increases the CP validation gradually becomes better. The reason for the improved matching of CP activity patterns at higher number of agents may be found in the fact that CP activity patterns are composed of activities of all agents at the CP. As such more agents may provide a more complete set of activities at the CP.

% Next, it can bee seen that the test set values for both agent and CP are higher than the training set. The logical reason for this increase is that the large uptake of both agents and deployment of CPs over time during the time of the training set \cite{cijferselekctrischvervoer}.  For agent validation values this implies that competition between agents in the real system could occur that is not present in the simulation. For CP validation both a growth in new agents with their specific activity patterns and the existence of new alternative CPs affect the CP validation value. Given these results and nuances we conclude that the model show a reasonable validation.

% % % % % % % % % % % % % % % % % % % % % % %
%               SIMULATION RESULTS 				 %
% % % % % % % % % % % % % % % % % % % % % % %

\section{Simulation Results}
\label{sec:SimResults}
While the focus of this research is primarily on the development of the agent based model, we do include some preliminary applications of the model to demonstrate how it might be used by policy makers. We have chosen applications that provide interesting findings that data analysis alone would not have been able to reveal. We conduct experiments that examine how increasing numbers of users impacts the overall system performance ~\cite{Helmus2018a}. A system that is used more intensely may evoke economies of scale or degradation of system performance due to increased competition.

\begin{figure}[ht!]
    \centering
    \includegraphics[width=\textwidth]{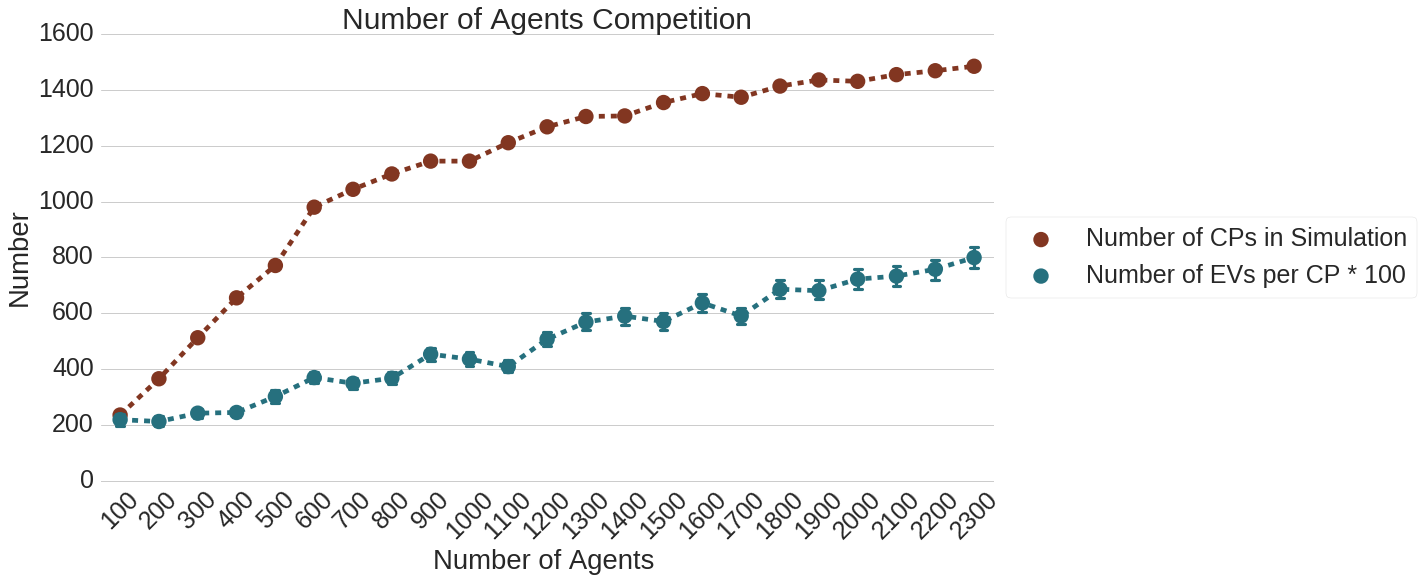}
    \caption{The number of CPs in the simulation and the number of agents per CP in the simulation for various numbers of agents in the simulation. Per value of the parameter we plot the mean numbers with the 95\% confidence interval.}
    \label{fig:competition number of agents}
\end{figure}

Figure \ref{fig:competition number of agents} shows how the number of CPs used by agents in the simulation and the number of unique users per CP vary as the number of users in the simulation is increased. It can be seen that while the number of EVs per CP increases almost linearly, the number of CPs used in the simulation initially steeply increases until 600 users and from there increases similarly to the numbers of unique users per CP. This indicates some economy of scale, where the ratio of the number of CPs divided by the number of EVs reduces. At 600 agents, there are 1000 CPs used and for 2300 agents there are only around 1500 CPs used. This is also demonstrated by the number of EVs per CP, initially at 200 agents each CP sees on average 2 different users (during the simulation), by 2300 agents each CP sees 8 unique users.

%%%%%%%%%%%%%%%%%%%%%%%%%%%%%%%%%%
%   COMPETITION NUMBER OF AGENTS
%%%%%%%%%%%%%%%%%%%%%%%%%%%%%%%%%%

\begin{figure}[ht!]
    \centering
    \includegraphics[width=0.9\textwidth]{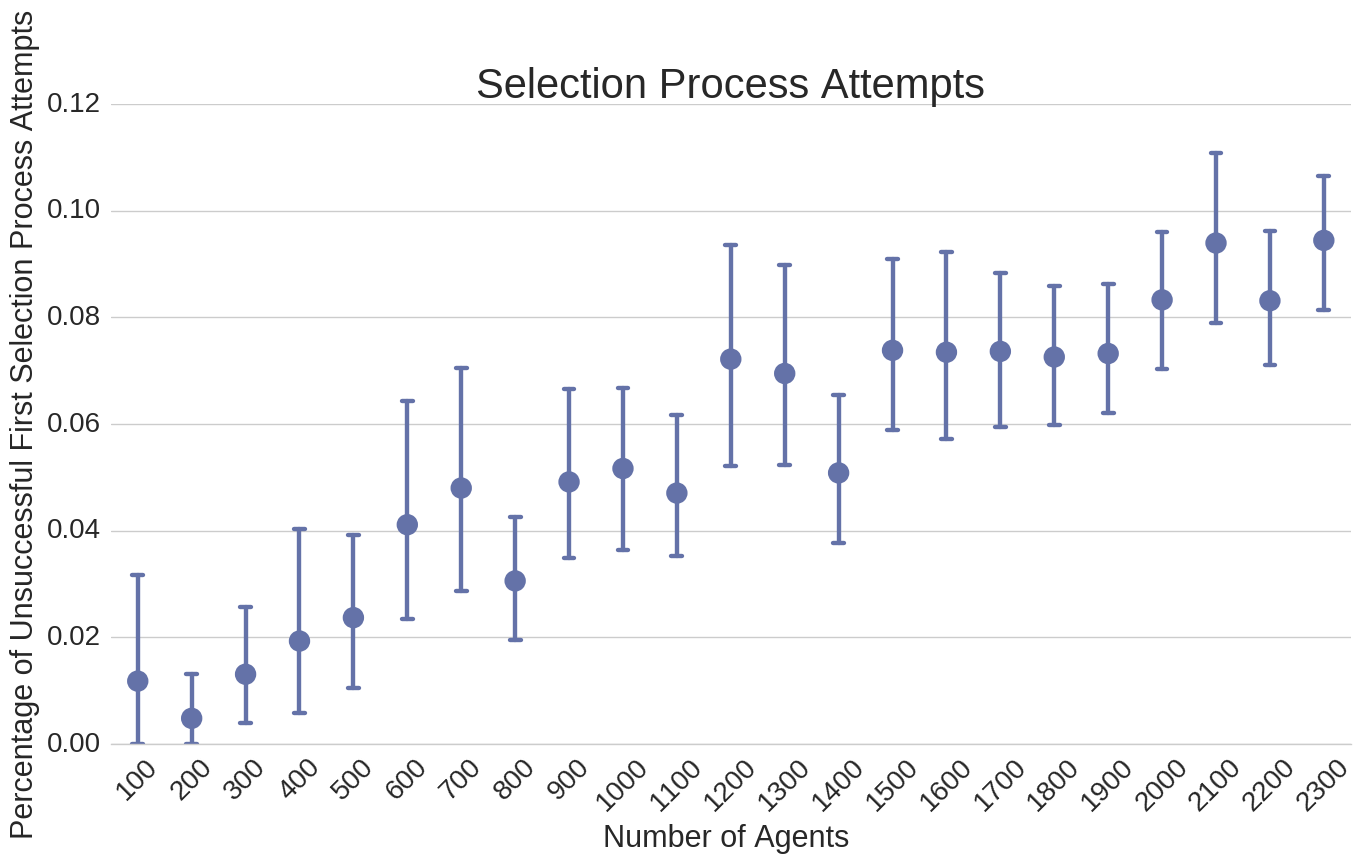}
    \caption{The percentage of failed (first) attempts in the selection process. Per value of the parameter we plot the mean percentages with the 95\% confidence interval.}
    \label{fig:percentage first fail}
\end{figure}

The competition in the EV system may be measured by the failed connection attempts of users at a CP. Note, these failed attempts are not part of transaction data as that only includes successful connection attempts \cite{Helmus2018a}. An unsuccessful connection attempt occurs when an agent has selected a cluster and tries to pick an occupied CP in that cluster. After the first attempt an agent may make consecutive attempts to connect, in these experiments we do not count this consecutive attempts and only count the first failed attempt. In Figure \ref{fig:percentage first fail} an analysis of the first failed attempts is shown in relation to an increasing number of users in the system. It can be seen that the number of failed attempts increases as more agents are present in the system, indicating that more competition is present. This demonstrates that within the current bounds of the system we do not see a significant change in the competition dynamics. We hypothesize that as the system grows in the future we may see a different regime of competition dynamics.

% % % % % % % % % % % % % % % % % % % % % % %
%               CONCLUSION                  %
% % % % % % % % % % % % % % % % % % % % % % %

\clearpage
\section{Conclusion}
\label{sec:conclusion}
%Note: the conclusion should be more extensive and include the (currently not available) results from the sensitivity analysis and validation.
In this paper we presented the SEVA model which enables simulation of charging behavior in the 4 largest cities of the Netherlands. The model is data-driven, making use of a rich dataset to extract charging behavior, and contains design concepts specifically for charging behavior based on literature findings. The internal workings of the model have been discussed in detail. We provided argumentation for the various choices of the parameters in the model using an sensitivity analysis. The model showed limited sensitivity towards agent validation, while CP validation showed more sensitivity. Lastly, we validated the model from the agent's point of view as well as from the CPs point of view and found that the model validates well.

The strength of the SEVA model is that it is data-driven and as such it can offer meaningful predictions. It provides a strong basis for further research on the assessment of charging infrastructure and the effects of infrastructure optimization methods. It enables artificial growth of both CP and EV users population in the system, which allows for testing rollout strategies and user adoption. The preliminary simulation results shows novel insights regarding system dynamics, such as economies of scale and competition. The economies of scale shows indications for non linear growth of the CP base given the growth of EV user. The competition effects relate to EV users not being able to charge at their preferred charging location which may be related to he perceived convenience of charging infrastructure \cite{Glombek2018}. As such, the SEVA model may be ale to provide insights on charging infrastructure rollout to policy makers that typical data analysis would not be able to reveal.

% % % % % % % % % % % % % % % % % % % % % % %
%               FUTURE WORK                 %
% % % % % % % % % % % % % % % % % % % % % % %

An Another strength of the model is its modularity which allows for adding new behavioral features or other system configurations. While the current model is focusing on connection activities, a logical extension of the model would be to include car type and transaction volume (in kWh) as well. The latter is part of the available data as well \cite{Mies}. Having this implemented typical scenarios for smart charging could be included in the simulation \cite{Lund2015}.

While the current model sets up individual behavior patterns in the model, a logical improvement - also for computational scale up - would be to focus on user types within the EV users population \cite{Helmus2015c}. From previous research it is known that there are more EV user types present in system. Having model that enables simulation of growth per user type allows to study the interactions between these user types which may be direct input for policy makers on what kind of potential EV users to stimulate.

\section*{References}

\bibliography{library} 

\section{Appendix}

\subsection{Overview of entities and their state variables}

\begin{table}[h]
    \centering
    \begin{tabularx}{\textwidth}{p{2.5cm}| p{3.5cm} X }
        \textbf{} & \textbf{State variables} & \textbf{Description} \\\hline
        \rowcolor{Gray}
        \textbf{Agent} & ID & Unique identifier. \\
        & Given charging transactions & List of charging transactions from the given data. \\\rowcolor{Gray}
        & Connected & Indicates if an agent is connected to a CP. \\
        & Time next activity & Date and time of the next activity. \\\rowcolor{Gray}
        & Active center & Location of the active cluster. Note that this is the center the agent is connected to or the center the agents plans to connect to next. \\
        & Active CP & ID of the active CP. Note that this is the CP the agent is connected to or `none' if the agent is disconnected. \\\hline\rowcolor{Gray}
        \textbf{Environment} & CP occupation & This variable stores whether each socket of each CP is occupied (and by which agent).\\
        & CP meta-data & Meta-data about each CP, for example the location (longitude, latitude) where the CP is located.\\\hline\rowcolor{Gray}
        \textbf{Simulation} & Agents & The agents contained in this simulation.  \\
        \textbf{handler}& Current time & The current time of the simulation.\\\rowcolor{Gray}
        & Sensors & Sensors to keep track of system metrics.\\\hline
    \end{tabularx}
    \caption{The entities of the model with their state variables.}
    \label{tab:entities}
\end{table}

\subsection{Model Parameters}
\label{sec:model parameters}

\textcolor{red}{tables should be altered to tech tables, currently screenshots from pdf in figures/AppendixA}

%\subsection{Process overview}
\label{subsection:process_overview}

\subsection{Model metrics}
\label{sec: model metrics}
\begin{table}[ht]
    \centering
    \begin{tabularx}{\textwidth}{l | p{4cm} X }
        \textbf{} & \textbf{Metric} & \textbf{Description} \\\hline
        \rowcolor{Gray}
        \textbf{Clustering} & Number of CPs & The total number of CPs in the simulation at which at least one agent in the system has had one or more charging transactions. \\
         & Number of clusters & The number of clusters for every agent in the simulation. \\
        \rowcolor{Gray}
         & Maximum distance ($d$)& For every agent in the simulation the maximum distance between a cluster and any of its CPs. \\
         & Walking preparedness ($w$) & The walking preparedness for every agent in the simulation, defined as the maximum distance plus 10\% with a minimum of the default value (see \ref{tab:parameters agent}). \\
        \rowcolor{Gray}
         & Number of CPs per cluster & The average number of CPs per cluster for every agent in the simulation.\\\hline
        \textbf{Competition} & Number of agents per CP & For every CP the number of agents that have been charging at that CP at least once during the simulation. \\
        \rowcolor{Gray}
        & Selection process attempts & For every agent in the simulation the consecutive failures and successes of the selection process, where a fail indicates that no CP within the desired cluster could be selected at the preferred time due to every CP being occupied.   \\\hline
        \textbf{Run time} & Run time per agent initialization & The run time for initializing a single agent, for every agent in the simulation. \\
        \rowcolor{Gray}
         & Run time per simulation & The run time for a complete simulation. \\\hline
        \textbf{Validation} & Validation error per agent & For every agent in the system the validation error (see Section \ref{sec:metrics} for details).\\
        \rowcolor{Gray}
        & Validation error per CP & For every CP the validation error (see Section \ref{sec:metrics} for details). \\\hline
    \end{tabularx}
    \caption{Output metrics of the simulation model.}
    \label{tab:metrics}
\end{table}

\subsection{Appendix for \ref{subsect:validation} - Typical validation values}
\label{Appendix validation}

\subsection{Appendix for \ref{sec:base_model_evaluation} - Sensitivity Analysis and Model Evaluation}
%%%%%%%%%%%%%%%%%%%%%%%%%%%%%%%%%%
%   NUMBER OF AGENTS
%%%%%%%%%%%%%%%%%%%%%%%%%%%%%%%%%%

\bigbreak
%%%%%%%%%%%%%%%%%%%%%%%%%%%%%%%%%%
%   SIMULATION REPEATS
%%%%%%%%%%%%%%%%%%%%%%%%%%%%%%%%%%

In order to select a default value for the number of simulation repeats simulations were performed (with 2000 agents) varying from 5 to 30 simulation repeats. The number of simulation repeats appeared to have limited effect on all validation values (see figure \ref{fig:validation simulation repeats} in the appendix). The reason for this limited effect is as follows. There are multiple activities for each agent in a single simulation (on average 200 transactions). These repeating activities and a large number of agents causes the need for less repeats of the experiments. The initialization, as well as the simulation itself, do contain randomness though, which is why we lower the number of repeats no lower than five.
\
\begin{figure}[ht!]
    \centering
    \includegraphics[width=\textwidth]{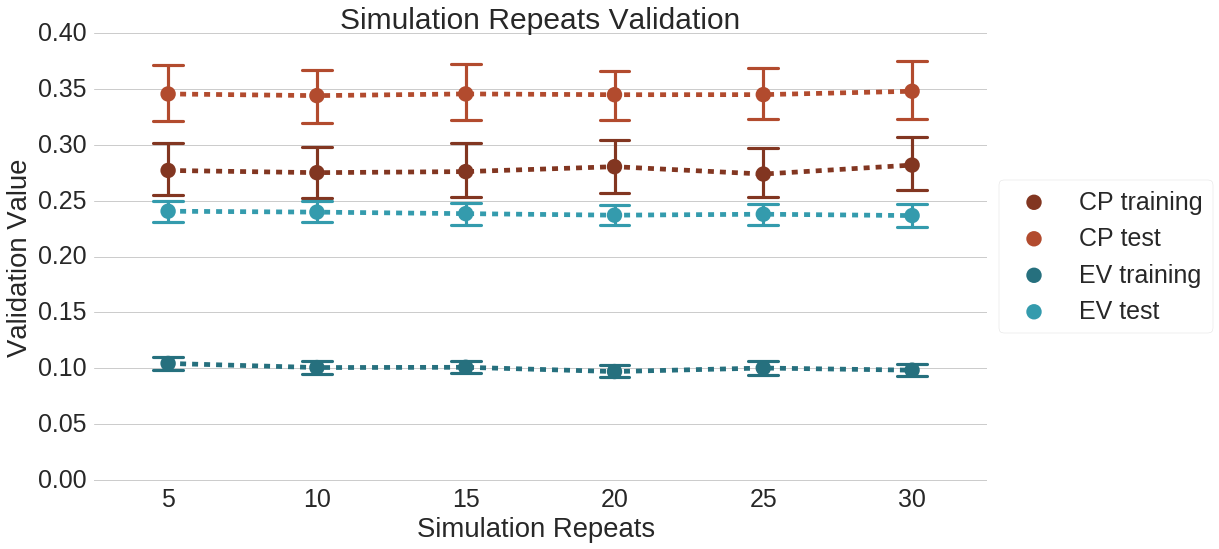}
    \caption{The results of the experiment where the number of simulation repeats is varied. Per value of the parameter we plot the mean validation values with the 95\% confidence interval.}
    \label{fig:validation simulation repeats}
\end{figure}

%%%%%%%%%%%%%%%%%%%%%%%%%%%%%%%%%%
%   WARMUP
%%%%%%%%%%%%%%%%%%%%%%%%%%%%%%%%%%
The last simulation setup parameter to be decided upon is the warmup period. This period is the length of time at the beginning of the simulation which is not taken into consideration during validation.
The warmup period appears to have little effect on the validation metrics. This can be explained by the fact that the initialization generates behavior and thus the agents are exhibiting proper behavior from the moment the simulation starts. In order avoid to be sure inconsistencies at the initialization, a warm-up period of 7 days was chosen.

\begin{figure}[ht!]
    \centering
    \includegraphics[width=\textwidth]{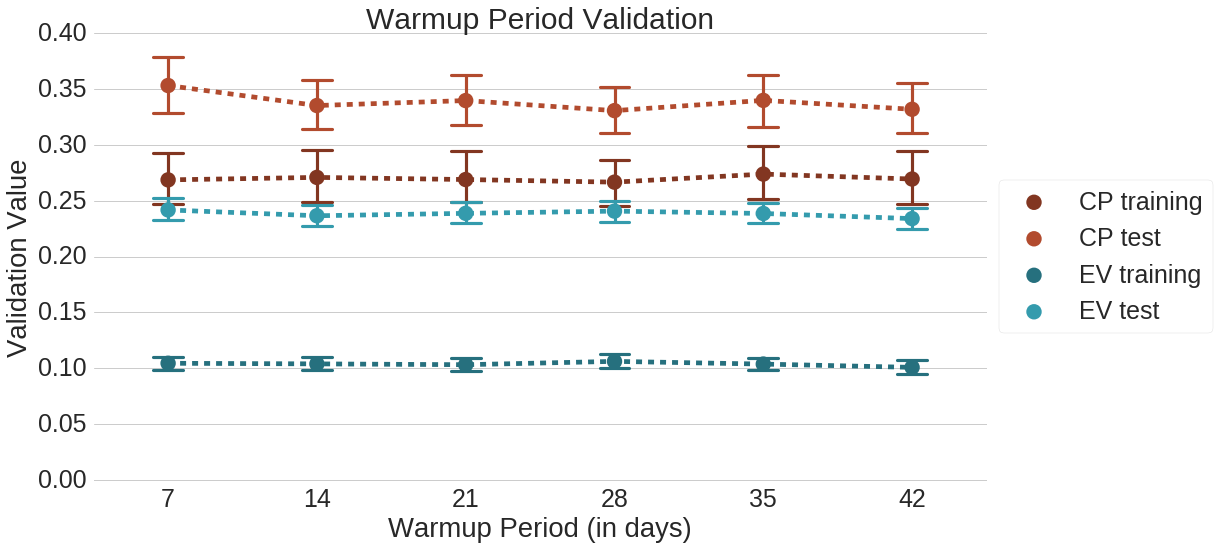}
    \caption{The results of the experiment that varies the warmup period and measures the validation values for the agents and CPs. Per value of the parameter we plot the mean validation values with the 95\% confidence interval.}
    \label{fig:validation warmup period}
\end{figure}

The BIRCH algorithm has a second input parameter, namely the BIRCH branching factor\cite{Zhang1996}, which controls the maximum number of entities in a cluster before a sub cluster is defined. Analysis of this parameter shows zero influence on the clustering. This is due to the fact that the maximum value of CPs within a cluster is not reached at the current state of charging infrastructure density.

%%%%%%%%%%%%%%%%%%%%%%%%%%%%%%%%%%
%   MINIMUM RADIUS
%%%%%%%%%%%%%%%%%%%%%%%%%%%%%%%%%%

The influence of this minimum radius parameter can be seen in \ref{fig:validation minimum radius}. As the minimum radius increases, the numbers of CPs to be taken into consideration increases as well. It is expected that this parameter mainly affects the CP validation as sessions may be distributed over more CPs, leading to CPs facilitating more unique users each with their own activity pattern. From the sensitivity analysis it appears that the minimum radius has little influence on the CP validation metric. A distance of $150$ meters was chosen as the default value. This means that whenever an agent has a cluster that has a radius of less than $150$ meters, it will still consider alternative CPs within a radius of $150$ meters (the default minimum radius).

%%%%%%%%%%%%%%%%%%%%%%%%%%%%%%%%%%
%   RETRY TIME
%%%%%%%%%%%%%%%%%%%%%%%%%%%%%%%%%%
The last parameter for the sensitivity analysis is the retry time of a cluster. Whenever all CPs within the cluster of an agent are occupied, the agent will try to connect again after this retry time. The retry time appeared to have little influence on both CP and EV validation. Large values though, larger than $120$ minutes are less preferred due potential skewed distributions, while low values, less than $20$minutes, would cause an large and unrealistic number of retries. Also, for to reduce computational complexity, a the default value of $20$ minutes, which is equal to the default bin size, was chosen.

\end{document}